| Category No. | TN929 | Public | Yes ☑ No ☐ |
|---|---|---|---|
| UDC | 621.39 | Thesis No. | D-10617-308-2021 |

# Chongqing University of Posts and Telecommunications

# Thesis for Master's Degree

| | |
|---|---|
| TITLE | **Fuzzy Based Secure Clustering Schemes for Wireless Sensor Networks** |
| COLLEGE | **School of Communication and Information Engineering** |
| AUTHOR | **Mohd Adnan** |
| STUDENT NUMBER | **L201820045** |
| DEGREE CATEGORY | **Master of Engineering** |
| MAJOR | **Information and Communication Engineering** |
| SUPERVISOR | **Professor Yang Tao** |
| SUBMISSION DATE | **2021/06/01** |

# DECLARATION OF ORIGINALITY

   I declare that this thesis/dissertation is the result of an independent research I have made under the supervision of my supervisor. It does not contain any published or unpublished works or research results by other individuals or institutions apart from those that have been referenced in the form of references or notes. All individuals and institutions that have made contributions to my research have been acknowledged in the Acknowledgements. In addition, I understand that any false claim in respect of this work will result in disciplinary action in accordance with regulations of Chongqing University of Posts and Telecommunications.

**Signature of author:** 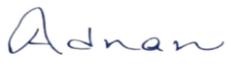            **Date:  2021/06/01**

# COPYRIGHT PERMISSION LETTER

   I hereby authorize Chongqing University of Posts and Telecommunications (CQUPT) to use my thesis to be submitted to the CQUPT, namely I grant an irrevocable and perpetual license that includes nonexclusive world rights for the reproduction, distribution, and storage of my thesis in both print and electronic formats to CQUPT and to the relevant governmental agencies or institutes to reproduce my thesis. I also extend this authorization to CQUPT, for the purposes of reproducing and distributing single micro form, digital, or printed copies of the thesis or dissertation on demand for scholarly uses.

**Signature of Student:** 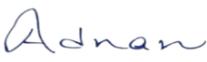          **Date:2021/06/01**

**Signature of Supervisor:** 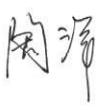        **Date:  2021/06/1**

# ABSTRACT

Due to many constraints, wireless sensor networks (WSNs) allow new solutions and necessitate non-traditional methodologies for protocol design. A proper balance between communication and signal/data processing capacities must be sought due to the need for low system complexity as well as low energy consumption. Since the last decade, this has motivated a major effort in research activities, standardization processes, and industrial investments in this area. Wireless Sensor Networks (WSNs) have gained widespread interest in recent years as a result of developments in the fields of information and communication technology, as well as the field of electronics. This cutting-edge sensing technology employs a large number of sensor nodes that are strategically placed in a given region to detect any constantly changing physical phenomenon. These tiny sensor nodes sense and process data before transmitting it to a base station or sink through a radio frequency (RF) channel. The small size of these sensors is advantageous because it allows them to be easily integrated inside any system or environment. This feature has prompted the use of WSNs in a wide range of applications, especially monitoring and tracking, with surveillance being the most prominent. However, the limited size of sensor nodes limits their resource capabilities. WSNs are typically deployed in applications where human interference is dangerous or difficult. The sensed data may be required to make critical decisions in emergency applications. As a result, maintaining network connectivity is critical. To prolong the network's lifespan, it is important to make the best use of the available resources. If any node loses control, the entire network connection fails, and the deployment will be rendered ineffective. As a result, the majority of WSN research has focused on energy efficiency, with the design of energy efficient routing protocols playing a key role.

This dissertation presents three independent novel approaches for distinct scenarios to solve one or more open challenges. The first concern explains the focus on the lifetime of the networks: this dissertation will utilize a fuzzy logic-based clustering protocol (unequal clustering) with multi-hop transmission for load balancing, energy consumption minimization, and network lifetime prolongation. The protocol forms unequal clusters with cluster head (CH) being selected by fuzzy logic with competition radius. Node distance to the base station, concentration, and residual energy are input variables. The second concern focuses on network stability: we design a type 2 fuzzy logic-based clustering schemes in a multi-hop WSN to reduce energy consumption and improve network scalability. In this clustering scheme, we propose a cluster head (CH) selection strategy where a sensor node is elected as a CH based on type 2 fuzzy logic inputs. To balance the load of CHs we also select their radius size based on the fuzzy logic inputs.

Finally, the third concern is focus on the utility of game theory in defensive Wireless Sensor Networks (WSN) from selfish nodes and malicious behavior. Game theory can effectively model WSNs malicious attacks because of their low complexity and scalability. The study, thus, explores different WSN defense strategies from both external attackers and internal nodes acting selfishly or maliciously using the game theory approach. Also, the chapter highlights the general trust model for decision-making using the game theory framework. Besides, the chapter demonstrates the significance of the theory in ensuring WSN security from acute attacks and its role in enhancing trustworthiness in data and cooperation of nodes in various WSN architectures.

# 摘要

由于存在许多限制，无线传感器网络（WSN）需要使用新的解决方案，并且 A 采用非传统方法进行协议设计。由于低系统复杂度和低能耗的需求，因此必须在通信和信号/数据处理能力之间寻求适当的平衡。近十年以来，这促使人们致力于该领域的研究活动，标准化流程和工业投资方面等。近年来，由于信息和通信技术以及电子领域的发展，无线传感器网络（WSN）引起了人们广泛的关注。这项先进的传感技术可以在给定区域中部署大量传感器节点来监测所有不断变化的外部环境。这些微小的传感器节点会先感测和处理数据，然后再通过射频（RF）信道将其传输到基站或接收器。这些小尺寸的传感器能够很容易地集成在任何系统或环境中。此特性促使 WSN 在各种应用中广泛应用，尤其是监测和跟踪，其中监测最为突出。但是，传感器节点的大小导致它们的资源能力有限。WSN 通常部署在危险或困难的人为干扰应用中。在紧急应用关键决策中可能会用到这些监测数据。因此，保持网络连接至关重要。为了延长网络的寿命，充分利用可用资源很重要。如果任何节点失去控制，将导致整个网络连接失败，并且部署将变得无效。因此，大多数 WSN 研究都集中在能效上，而节能路由协议的设计起着关键作用。

本文针对不同的场景提出了三种独立的新方法，以解决一个或多个上述的公开挑战。第一项工作针对网络寿命的研究：本论文将利用基于模糊逻辑的群集协议（非均匀群集）和多跳传输来实现负载平衡，能耗最小化和网络生存期的延长。该协议形成非均匀的簇，簇头（CH）由具有竞争半径的模糊逻辑选择。输入变量为到基站的节点距离，节点密度和剩余能量。第二项工作是网络稳定性：我们在多跳 WSN 中设计了一种基于 2 型模糊逻辑的聚类方案，以减少能耗并提高网络可扩展性。在此聚类方案中，我们提出了一种簇头（CH）选择策略，其中基于 2 类模糊逻辑输入将传感器节点选举为 CH。为了平衡 CH 的负载，我们还根据模糊逻辑的输入选择其半径大小。最后，第三项工作是博弈论在无线传感器网络（WSN）

中防御自私节点和恶意行为的应用。由于博弈论的低复杂性和可扩展性，可以有效地对 WSN 恶意攻击进行建模。因此，该研究使用博弈论方法探索了来自外部攻击者和内部节点的自私或恶意行为的不同 WSN 防御策略。此外，本文重点介绍了基于博弈论框架的一般决策信任模型。此外，本文还验证了博弈论在确保 WSN 免受急性攻击的安全方面的重要性，以及在增强各种 WSN 架构中的节点数据和节点协作中的可信度方面的作用。

关键字：WSN，2 型模糊逻辑，博弈论，模糊逻辑，网络寿命，可伸缩性，能耗，多跳通信。

# CONTENTS

# List of Figures

# List of Tables

XII

# Abbreviations

| | |
|---|---|
| ADCs | Analog to Digital Converters |
| BS | Base Station |
| CH | Cluster Head |
| CH-MSG | Cluster Head Message |
| CM | Cluster Member |
| COA | Center of Area |
| DoS | Denial of Service |
| DS | Distance |
| FHSS | Frequency Hopping Spread Spectrum |
| FIS | Fuzzy Interference System |
| FL | Fuzzy Logic |
| FND | First Node Died |
| HND | Half Node Died |
| LMAC | Lightweight medium access |
| MCU | Microcontroller Unit, |
| PCH | Prime Cluster Head |
| RE | Residual Energy |
| SNR | Signal-to-Noise Ratio |
| TH | Threshold |
| WSN | Wireless Sensor Networks |

# Chapter 1 Introduction

The Wireless Sensor Network (WSN) is a network of spatially dispersed and dedicated sensor nodes that track and record the physical conditions of the environment while organizing the data collected at a central location. WSNs compute environmental variables such as temperature, vibration, pollution levels, humidity, wind, and so on. The research group focused on wireless sensor networks has created a variety of essential frameworks, algorithms, and abstractions. Wireless sensor networks are designed to last for a long time. Since wireless sensors typically use batteries, having a long life cycle reduces the power consumption of individual nodes. As a result, numerous power-saving mechanisms have been developed, implemented, analyzed, and tested in both simulators and real-world deployments. Many of them are specifically applicable to intelligent objects. Wireless sensor networks are described similarly to smart objects, and much of the development of smart objects has occurred in the community surrounding wireless sensor networks. Wireless sensor networks are made up of small nodes equipped with a wireless communication interface that self-configure into networks from which sensor readings can be transmitted.

A sink or base station serves as a point of communication between users and the network. By inserting queries and collecting results from the sink, one can retrieve necessary information from the network. A wireless sensor network usually consists of hundreds of thousands of sensor nodes. The sensor nodes can communicate with one another through radio signals. A wireless sensor node contains sensing and computing equipment, as well as radio transceivers and power components. Individual nodes in a wireless sensor network (WSN) are resource constrained by design, with limited processing speed, storage space, and communication bandwidth. Following deployment, the sensor nodes are responsible for self-organizing an effective network infrastructure, often with multi-hop contact with them. The onboard sensors then begin collecting relevant data. Wireless sensor devices often respond to commands from a "control site" to carry out specific instructions or provide sensing samples. The sensor nodes can operate in either a continuous or event-driven mode Wireless sensor systems can be outfitted with actuators to "act" in response to certain conditions. As defined in [1] these networks are often referred to as Wireless Sensor and Actuator Networks. Due to many constraints, wireless sensor networks (WSNs) allow new applications and necessitate non-traditional paradigms for protocol design. A proper balance between communication and signal/data processing capacities must be sought due to the need for low system complexity as well as low energy consumption (i.e. long network lifetime). This motivates a huge effort in research activities, standardization process, and industrial investments on this field since the last decade [2].

## 1.1 Design Issue of Wireless Sensor Networks

The deployment of sensor networks, which are a subset of wireless ad hoc networks, introduces a slew of new challenges. Sensor nodes communicate over lossy wireless lines with no infrastructure. An additional challenge is the sensor nodes' small, typically non-renewable energy supply. To optimize the network's lifespan, the protocols must be built from the start with the goal of efficient energy resource management in mind. Wireless Sensor Network Architecture concerns are addressed in [3][4][5][6], and various platforms for simulation and testing of routing protocols for WSNs are discussed. Let us now go through each individual design problem in greater depth.

### 1.1.1. Fault Tolerance

Sensor nodes are highly unstable and are often installed in hazardous environments. Nodes can fail due to hardware issues, physical damage, or running out of energy. We anticipate that node failures would be much higher than in wired or infrastructure-based wireless networks. A sensor network's protocols should be able to detect these failures as soon as possible and be robust enough to manage a reasonably large number of failures while preserving overall network functionality. This is particularly critical in the design of the routing protocol, which must ensure that alternative paths are available for packet rerouting. Different implementation environments necessitate different levels of fault tolerance.

## 1.1.2. Scalability

Sensor networks range in size from a few nodes to potentially hundreds of thousands. Furthermore, the deployment density varies. For high-resolution data collection, node density can reach the point where a node has several thousand neighbors within its transmission range. The protocols used in sensor networks must be scalable to these levels while maintaining adequate efficiency.

## 1.1.3. Production Costs

Since several deployment models regard sensor nodes as disposable instruments, sensor networks may compete with conventional information gathering approaches only if individual sensor nodes can be generated at a low cost. The optimal target price for a sensor node should be less than $1.

## 1.1.4. Hardware Constraints

Any sensor node must have at least one sensing unit, one processing unit, one transmission unit, and one power supply. To allow location-aware routing, the nodes can have many built-in sensors or external devices such as a localization system. However, each additional feature comes at a cost, increasing the node's power consumption and physical size. As a result, added functionality must always be balanced against cost and low-power requirements.

## 1.1.5. Sensor Network Topology

WSNs have advanced in several respects, but they remain networks with limited resources in terms of energy, processing power, memory, and communications capabilities. Energy consumption is the most important of these constraints, as shown by the vast number of algorithms, techniques, and protocols built to save energy and thus prolong the network's lifespan. One of the most critical problems being researched to reduce energy consumption in wireless sensor networks is topology maintenance.

## 1.1.6. Transmission Media

Communication between nodes is typically accomplished through radio communication over common ISM bands. Some sensor networks, on the other hand, use optical or infrared communication, the latter of which has the advantage of being stable and practically interference free.

## 1.1.7. Power Consumption

Many of the problems of sensor networks, as we've seen, revolve around their limited power resources. The battery's size is limited by the size of the nodes. The issues of effective energy usage must be carefully considered in the design of software and hardware. Data compression, for example, can reduce the amount of energy required for radio transmission but requires additional energy for computation and/or filtering. The energy policy is often determined by the application; in some cases, it might be acceptable to switch off a subset of nodes in order to save energy, while in others, all nodes must be operational at the same time.

## 1.2.    Structure of Wireless Sensor Networks

A wireless sensor network's framework involves numerous topologies for radio communications networks. The following is a brief discussion of the network topologies that relate to wireless sensor networks.

## 1.2.1 Star Network (single point-to-multipoint)

A star network is a communications topology in which a single base station can send and/or receive messages from multiple remote nodes [7]. The remote nodes are not permitted to communicate with one another. The benefits of this network type for wireless sensor networks include simplicity and the ability to keep the remote node's power consumption to a minimum. It also enables low-latency communication between the remote node and the base station. The disadvantage of such a network is that the base station must be within radio transmission range of all individual nodes and is less robust than other networks due to its reliance on a single node to manage the network.

## 1.2.2. Mesh Network

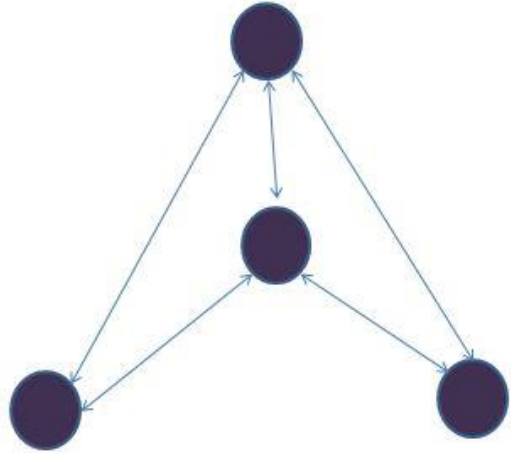

Fig. 1.1 Mesh network topology

A mesh network allows data to be transmitted from one node to another within the network's radio transmission range are shown in fig. 1.1. This enables multi-hop communications, which means that if a node wants to send a message to another node that is out of radio communications range, it can use an intermediate node to forward the message to the desired node. This network topology provides redundancy and scalability. If a single node fails, a remote node can still communicate with any other node within its range, which can then forward the message to the desired location. Furthermore, the network's range is not necessarily limited by the distance between single nodes; it can simply be extended by adding more nodes to the system. The disadvantage of this type of network is that the power consumption for nodes that implement multi-hop communications is generally higher than for nodes that do not have this capability, which frequently limits battery life. Furthermore, as the number of communication hops to a destination increases, so does the time to deliver the message, especially if low power operation of the nodes is required.

### 1.2.3. Hybrid star – Mesh Network

A hybrid of the star and mesh networks provides a robust and versatile communications network while keeping wireless sensor node power consumption to a minimum. The sensor nodes with the lowest power in this network topology are not enabled to forward messages. This ensures that power consumption is kept to a minimum.

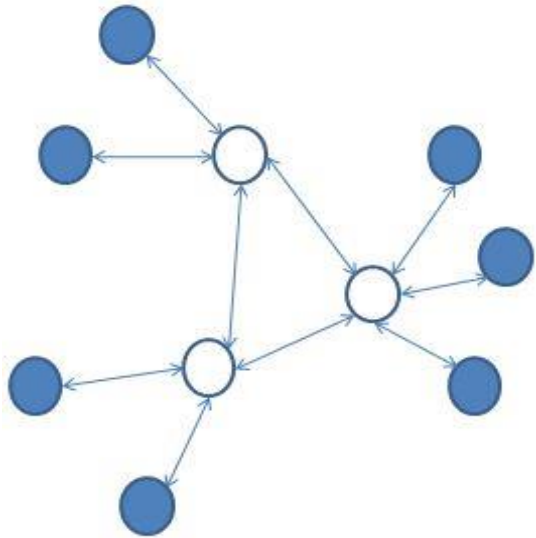

Fig. 1.2 Hybrid Star Mesh Network Topology

The transmissions are shown in fig. 1.2. Other network nodes, on the other hand, have multi-hop capability, allowing them to forward messages from the low power nodes to other network nodes. In general, nodes with multi-hop capability are more powerful and, if possible, are plugged into the electrical mains line. This is the topology implemented by the up-and-coming mesh networking standard known as ZigBee.

### 1.3.  Structure of Wireless Sensor Nodes

A sensor node is made up of four main parts: a sensing unit, a processing unit, a transceiver unit, and a power unit. It also has application-specific features like a position finding system, a power generator, and a mobilizer. The two main components of sensing units are sensors and analog to digital converters (ADCs). The ADC converts analogue signals from sensors into digital signals, which are then fed into the processing unit. The processing unit is typically associated with a small storage unit, and it is responsible for the procedures that allow the sensor node to collaborate with the other nodes to complete the assigned sensing tasks. A transceiver device connects the node to the network. A sensor node's power unit is an essential component. Power units can be supported by a power scavenging device, such as solar cells. The other subunits of the node are application- based.

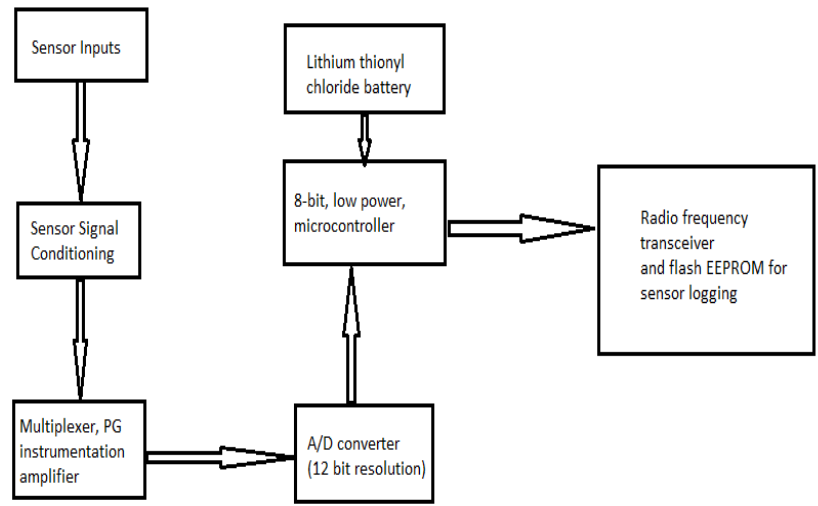

Fig. 1.3 Block diagram of sensor node

A functional block diagram of a flexible wireless sensing node is shown in Fig. 1.3. The modular design approach provides a flexible and scalable foundation for meeting the needs of a wide range of applications. Depending on the sensors to be deployed, the signal conditioning block, for example, can be reprogrammed or replaced. As a result, the wireless sensing node can work with a wide variety of sensors. Similarly, the radio link can be turned off as needed to meet the wireless range requirements of a given application as well as the requirement for bidirectional communications. The remote nodes acquire data using flash memory in response to a command from a base station or an event sensed by one or more node inputs. Furthermore, the embedded firmware can be upgraded in the field through the wireless network.

The microprocessor has a number of functions including:

1. Managing data collection from the sensors
2. Performing power management functions

3. Interfacing the sensor data to the physical radio layer
4. Managing the radio network protocol

It is critical for any wireless sensing node to use as little power as possible. Typically, the radio subsystem necessitates the greatest amount of power. As a result, data is only sent over the radio network when it is required. An algorithm will be loaded into the node based on the sensed case to determine when to submit data. It is also critical to keep the sensor's power consumption to a minimum. As a result, the hardware should be designed so that the microprocessor can precisely monitor power to the antenna, sensor, and sensor signal conditioner [3].

## 1.4. Communication Structure of Wireless Sensor Networks

Sensor nodes are typically distributed throughout a sensor area. Each of these dispersed sensor nodes can collect data and route it back to the sink and end users. Data is routed back to the end user via the sink in a multi-hop infrastructure-less architecture. The sink and task manager nodes can communicate via the Internet or satellite. Fig. 1.4 depicts the protocol stack used by the sink and sensor nodes. This protocol stack combines power and routing knowledge, integrates data with networking protocols, efficiently communicates power over the wireless medium, and promotes sensor node collaboration. The protocol stack is composed of the following layers: application layer, transport layer, network layer, data link layer, physical layer, power management plane, mobility management plane, and task management plane. Furthermore, the power, mobility, and task management planes keep track of power, movement, and task distribution among sensor nodes. These planes aid sensor nodes in coordinating sensing tasks and reducing overall energy consumption.

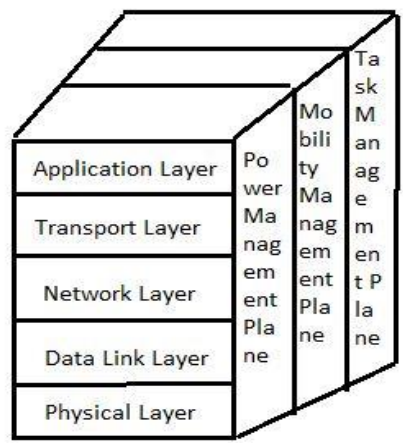

Fig. 1.4 WSN protocol stack

## 1.5. Energy Consumption Issue in Wireless Sensor Network

Because sensor nodes are typically powered by batteries, energy consumption is the most important factor in determining the life of a sensor network. Energy optimization can be more difficult in sensor networks because it involves not only reducing energy consumption but also extending the network's life as much as possible. The optimization can be accomplished by incorporating energy awareness into all aspects of design and operation. This ensures that energy awareness is implemented in groups of communicating sensor nodes as well as the entire network, rather than just individual nodes [8].

A sensor node usually consists of four sub-systems:

1. Computing subsystem: It is made up of a microprocessor (microcontroller unit, MCU) that controls the sensors and implements communication protocols. MCUs typically operate in a variety of modes for power management. Because these operating modes require power, the energy consumption levels of the various modes should be considered when calculating the battery lifetime of each node.
2. Communication subsystem: It is made up of a short-range radio that communicates with neighboring nodes as well as the outside world. Radios can operate in a variety of modes. When not transmitting or receiving, it is critical to completely shut down the radio rather than putting it in idle mode to save power.

3. Sensing subsystem: It is made up of sensors and actuators that connect the node to the outside world. Energy consumption can be reduced by using low-power components and saving power at the expense of performance that is unnecessary.

4. Power supply subsystem: It is made up of a battery that powers the node. The amount of power drawn from a battery should be checked because if a high current is drawn from a battery for an extended period of time, the battery will die faster even if it could have lasted longer. Typically, the rated current capacity of a sensor node battery is less than the minimum energy consumption. A battery's lifetime can be extended by drastically reducing its current or even turning it off frequently. Various protocols and algorithms have been studied all over the world to reduce the overall energy consumption of the sensor network. The lifetime of a sensor network can be significantly extended if the operating system, application layer, and network protocols are designed to be energy conscious. These protocols and algorithms must be aware of the hardware and capable of utilizing special features of microprocessors and transceivers to reduce the energy consumption of the sensor node. This may encourage the development of a custom solution for various types of sensor node design. Different types of sensor nodes deployed result in various types of sensor networks.

## 1.6. Security Issue in Wireless Sensor Networks

Security issues in sensor networks depend on the need to know what we are going to protect. The authors of [9] identified four security goals for sensor networks: confidentiality, integrity, authentication, and availability. [10] Introduces another security goal for sensor networks. Confidentiality refers to the right to hide a message from a passive intruder, through which the message transmitted on sensor networks remains private. The ability to ensure that a message has not been tampered with, altered, or modified while it was on the network is referred to as integrity. Verification it is necessary to determine the authenticity of the message's origin by deciding if the messages are from the node from which they claim to be. The availability of a node determines whether or not it has the opportunity to use the services and whether or not the network is accessible for the messages to pass on. Freshness denotes that the receiver receives new and fresh data while ensuring that no adversary can replay old data. This is particularly critical when WSN nodes use shared-keys for message communication, as a possible adversary may launch a replay attack using the old key while the new key is being refreshed and propagated to all WSN nodes [11].To achieve the freshness the mechanism like nonce or time stamp should add to each data packet.

After establishing a base of security objectives in sensor networks, identify the main potential security threats in sensor networks [12]. Attacks on routing loops target the information shared between nodes. When an attacker modifies and replays the routing information, false error messages are produced. Routing loops attract or repel network traffic while increasing node-to-node latency. The selective forwarding attack affects network traffic by assuming that all network nodes are trustworthy in forwarding messages. Malicious nodes in a selective forwarding attack actually drop those messages rather than forwarding all messages. When a malicious node cherry picks messages, it decreases latency and deceives neighboring nodes into thinking they are on a shorter path. The effectiveness of this attack is determined by two factors. First, the malicious node's location; the closer it is to the base stations, the more traffic it will draw. The second factor is the number of messages that are dropped. As a selective forwarder loses more messages and forwards less, it maintains its energy level, allowing it to continue to deceive neighboring nodes. The adversary draws traffic to a compromised node in sinkhole attacks. The best way to create a sinkhole is to position a malicious node where it can draw the majority of the traffic, likely closer to the base station or as a malicious node deceptively masquerading as a base station.

One explanation for sinkhole attacks is to allow selective forwarding in order to direct traffic to a compromised node. Because of the design of sensor networks, where all traffic flows to a single base station, this form of attack is more vulnerable. Sybil attacks are a form of attack in which a node generates multiple illegitimate identities in sensor networks by forging or stealing valid node identities. Sybil attacks can be used against routing algorithms and topology management, reducing the efficacy of fault tolerant schemes like distributed storage and disparity. Another malicious aspect is regional routing, which allows a Sybil node to appear in several locations at the same time. Wormhole attacks enable an adversary closer to the base station to fully interrupt traffic by tunneling messages over a low latency connection. In this case, an attacker convinces nodes that are multi-hop away that they are closer to the base station. This results in the creation of a sinkhole because the opponent on the other side of the sinkhole has a stronger path to the base station. In Hello flood attacks, a Broadcasted message with higher transmitting power impersonates the HELLO message from the base station. Message receiving nodes presume that the HELLO message sending node is the nearest and attempt to route all of their messages through it. During this form of attack, all nodes will react to HELLO floods, wasting energy. The real base station will also broadcast similar messages, but only a few nodes will respond to it. At the physical level, denial of service (DoS) attacks trigger radio jamming, network protocol interference, battery depletion, and other problems. A particular form of DoS attack, Denial-of-service attack, has been investigated in [13], in which the power supply of a sensor node is attacked. This form of attack can reduce sensor lifetime from years to days, wreaking havoc on a sensor network.

The layering-based security approach are as follows:

a) **Application layer:** Since data is collected and maintained at the application layer, it is critical to ensure data reliability. [14] has proposed a resilient aggregation scheme for a cluster-based network in which a cluster leader serves as an aggregator in sensor networks. This strategy, however, is only valid if the aggregating node is within range of all the source nodes and there is no intervening aggregator between the aggregator and the source nodes. Cluster leaders use cryptographic techniques to ensure data reliability to prove the validity of the aggregation.

b) **Network layer:** Network layer is responsible for routing of messages from node to node, node to cluster leader, cluster leaders to cluster leaders, cluster leaders to the base station and vice versa.

c) **Data link layer:** The data link layer is in charge of error detection and correction, as well as data encoding. The connection layer is vulnerable to jamming and denial-of-service attacks. Tiny Sec [15] has developed link layer encryption that is based on a key management scheme. An attacker with higher energy efficiency, on the other hand, can still rage an attack. Protocols with better anti-jamming properties, such as LMAC, are a viable counter measure at this layer.

d) **Physical Layer:** The physical layer emphasizes on the transmission media between sending and receiving nodes, the data rate, signal strength, frequency types are also addressed in this layer. Ideally FHSS frequency hopping spread spectrum is used in sensor networks.

# Chapter 2 Unequal Clustered Routing Protocol for Wireless Sensor Networks

## 2.1 Chapter Overview

Due to their wide usage in various applications such as disaster management, health, and home applications, wireless sensor networks (WSNs) have amassed a lot of attention. Developments such as micro-electrical-mechanical, wireless communication, and digital systems have allowed the use of miniaturized wireless sensors. A WSN consists of several miniature detector nodes. These nodes are deterministically/ randomly deployed to monitor specific target areas [3]. During sensing, processing, and transmitting the collected data, the sensor nodes exhaust a lot of energy. These nodes are supplied with a low-powered battery that is unique but rechargeable. The nodes' energy needs to be sparingly used so as long the life of the network is attained. One of the finest ways to decrease energy consumption by the network is grouping the sensor nodes in a certain characteristic that forms a cluster. Each cluster has a cluster head (CH) that allows communication with a base station (BS). The CHs gather data from the other member nodes of their specific cluster [16]. CH-BS communication is made up of either multi-hop or single-hop communication models. In the latter model, a CH communicates directly to a BS. On the other hand, multi-hop communication is characterized by a CH sending its data through other CHs to get to the BS eventually [17].

The approach is found to be logical in prolonging the network lifetime of WSNs. Also, multi-hop communication in a cluster reduces the number of links, thus avoiding congestion. Besides, multi-hop communication can allow member nodes to assist the CH to share data fusion, thus efficiently decreasing the energy consumption by the CHs. This will help in the extension of the lifetime of the networks [18]. However, in multi-hop communication, the CHs close to the BS is characterized by massive data trafficking. This is known as an energy hot spot [19]. To avoid an energy hole problem, unequal clustering algorithms can be utilized. The algorithms include keeping the clusters in the proximity of the BS trivial in size than the clusters away. This ensures CHs near the BS are more ready to allot the data traffic. The scalable nature of WSNs is also improved. Studies on clustering approaches in WSNs has mainly focused on establishing distributed and centralized protocols in computing groups of CHs. Centralized approaches are, however, inefficient in large-scale networks because collecting the whole amount of crucial details at the central BS is energy and time-consuming [20]. Distributed protocols are further efficient in large-scale networks. Here, a node can decide to emulate a CH or unite with an existing cluster. The process is determined by the nature of details obtained from adjacent nodes. Several dispersed clustering protocols have been suggested in different literature [21][22][23][24]. For [21], a majority of the protocols are either in the case of the iterative or probabilistic state. Probabilistic protocols are characterized by the probabilistic determination of a decision to become a CH [23][24]. Iterative protocols include an iterative process performed on the nodes to decide the becoming of a CH. Differently, clustering protocols can be contemplated as being static or dynamic. The clusters are permanent in static clustering. In dynamic clustering, the protocol is split into different rounds [22]. The clusters are then devised again in the next round. The additional overhead comes about when the repeated cluster is formed on the system. Some protocols take full advantage of fuzzy logic [25]. Even with incomplete information on environmental factors, real-time decisions are possible in fuzzy logic. Combining various environmental features based on preset rules and eventually coming up with a decision according to the finding is another important use of fuzzy logic.

In turn, fuzzy clustering algorithms use the same logic for combining different clustering frameworks to choose CHs. Fuzzy logic is a structure that functions almost similar to human logic. Fuzzy logic is made up of 4 main components, including fuzzifier, defuzzifier, fuzzy rule, and inference engines [26].

(a) Fuzzifier: This component maps every input value into the correlating fuzzy set. It assigns a degree of membership or truth value to each set.

(b) Inference engine: This component processes fuzzified values. It includes a variety of methods used to deduce the rules and a rule base.

(c) Fuzzy rule: This component is a sequence of IF-THEN directives where the output fuzzy variables are related to the input linguistic variables.

(d) Defuzzifier: Defuzzification is done by this component, where a solution space is mapped into one crisp input value.

The distance decides the inequality in cluster formation between the sink and the CHs. Therefore, CHs near the sink is smaller than the CHs away from the sink.

Major contribution are as follows:

This chapter explain fuzzy logic as the algorithm of choice. The algorithm includes unequally-sized clusters, where their cluster radius is determined through a distributed fuzzy logic approach. To minimize energy consumption, we use multi-hop transmission in the wsn with unequal clustering. Intelligence techniques of computation, such as particle swarm optimization [28], ant colony optimization [27], fuzzy logic, and genetic algorithms [29], have previously been used to solve several WSN issues. Here, we utilize three fuzzy variables- node's residual energy, concentration, and base station distance for computation of competition radius and choices CHs. The emergence of lots of uncertainties will include the use of the fuzzy logic approach [30][31]. To improve the chance of cluster radius and the chance of cluster head selection, we use one new variable, node concentration. It is important to consider this node concentration because it ameliorates the fuzzy logic algorithm, hence prolonging the WSN network lifetime. The multi-hop hot-spot problem is familiar in WSNs. The nodes close to the sink expire quickly because of the inter-cluster congestion. Reducing such a problem includes an approach known as unequal clustering.

## 2.2 Background and Related Works

In designing the WSN, energy is the most obligatory resource, so its battery life confines the sensor nodes' lifetime. The reduction in the energy consumption of nodes can lead to better network life. Several routing protocols have been developed to enhance network performance and lifetime. In this part, some famous clustered based routing protocols are explained. We have divided them into two categories. In the first category, few protocols where cluster heads are elected in a probabilistic manner are discussed. In the second category, some of the fuzzy logic clustering-based protocols are discussed.

### 2.2.1. Hierarchical Routing Protocols Based on Clustering

LEACH is a clustering algorithm, where the sensor nodes are probabilistically selected as viable CH in WSNs [32]. Each node selects an arbitrary number. A node automatically becomes the CH if the selected number is lower than the set threshold. After the CHs are determined, based on their distance, non-CH nodes join the CHs. The performance of the determined CH is initially low. Also, there is no consideration of residual energy during the selection of a CH.

PEGASIS [33] is a protocol that is chain-based and near optical and an upgrade of LEACH. In this protocol, each node communicates with an adjacent node. The nodes alternate in transmitting signals to the BS. This process reduces the quantity of energy eventually spent with each round. One major problem of this protocol includes the CHs directly sending data to the BS. The EAMMH protocol was started by inducing the attributes of multi-hop intra-clustering and energy-aware routing [34]. The utilization of the EAMMH protocol is divided into rounds. Each round starts with a launching phase where there is an organization of clusters. The initial phase is then followed by a steady-state phase, which includes data transfer to the BS. The second phase involves organizing the sensor nodes into clusters, forming a multi-hop intra-cluster communication. Multiple paths are established from every node to the CH. This provides energy-aware heuristic connectivity where the optimal path is chosen. Therefore, this protocol chooses CHs regarding their residual energy, especially when the survival time of the network is short.

In [35], the authors proposed an enhanced routing-Gi protocol for a mobile sink in WSNs to enhance the energy efficiency, minimize the packet loss rate and maximize the network lifetime. By optimizing the distance between the mobile sink and each CH at a specific time at a specific location, this proposed protocol conserves energy of CHs. At each round, based on the ratio between the energy level of CH and grid numbers, the mobile sink moves in a predetermined trajectory.

### 2.2.2. FL Based Clustering Protocol

Multiple studies have discussed Fuzzy Logic (FL) and how it is usable in clustering, especially in enhancing energy consumption. Some associated protocols will be discussed as follows:

In Hybrid Energy-Efficient Distributed Clustering (HEED) protocol, the primary parameter for selecting CHs is nodal residual energy. The process involves probabilistic election [21]. The distance between the node and BS is utilized to decide the CH whenever there is was a tie between two nodes. Experiments have been used to evaluate this protocol. They have demonstrated that data aggregation and clustering can, at least, double the duration of a WSN.

In a fuzzy clustering protocol that was proposed by Gupta et al. [25], the CHs are chosen at the BS. Input subjected to the system is characterized by residual energy, node centrality, and degree. In each round of selection, each node passes on its clustering output to the BS. This approach is incompatible with large-scale networks, such as WSN, because it suffers from scalability offsets. CHEF is another protocol similar to fuzzy clustering [22]. The BS does not have to gather clustering outputs from all nodes. This is because it selects CH in a diffused manner. A probabilistic process selects the provisional CHs. Fuzzy logic is then used to narrow the selection to the eventual CHs. Inputs administered to the fuzzy system are characterized by residual energy. The CHEF protocol provides a greater duration of network use. However, different from other protocols such as LEACH, CHEF may have non-uniformly distributed CHs in the network. This is mainly because of the probabilistic selection of the provisional CHs. In [36], this paper explains improved low-energy adaptive clustering hierarchy (LEACH) protocol for mobile sensor networks, which prolong the network lifetime and using fuzzy logic reduce the packet loss in a mobile sensing environment.

Recently, some unequal clustering protocols have been developed. These protocols are mainly based on creating smaller clusters around the sink. A significant distance to the BS also characterizes the smaller clusters. Sensor nodes near the sink are in smaller cluster sizes than other nodes located far away. EEUC protocol includes diffused unequal clustering in selecting CH through local competition [26]. A competitive radius is assigned to each node. This radius becomes more trivial when the nodes near the BS. The EEUC protocol, therefore, is characterized by unequal clustering. This protocol cannot support the processing load. The authors Bagci and Yazici in EAUCF protocol recommend unequal clustering algorithm [37]. In that study, fuzzy logic was used to process the competition range by considering the node's distance to the BS and residual energy. This protocol increases the network duration and solves the energy holes problem. The main offset here is the increased energy depletion at the CHs. In Two-Tier Distributed Fuzzy Logic-Based Protocol (TTDFP) [38], the authors proposed enhancing the data aggregation efficiency in the two-tier sensor networks. At first, optimum CH was selected based on probabilistic models. TTDFP utilizes the optimization framework to tune the two parameters in this tier, which are the threshold radius and the maximum competition radius, rather than the use of a trial-and-error approach to find the right blend of these parameters. In the second tier, fuzzy sets enhance the routing performance. In clustering, TTDFP uses three linguistic parameters, distance to BS, node energy, and node connectivity. In this, the residual energy. In this situation, using relative distance and average connection, the residual energy was resolved, and elected CHs based on the availability of node connectivity, which has ignored the energy in the networks.

MOFCA [39] is an unequal clustering protocol with multiple objectives. This protocol utilizes three parameters: node's residual energy, density, and distance to the sink. Also, the provisional CHs are selected through the use of fuzzy logic. Different from the other protocol, the competition radius in this algorithm is calculated. In [40], the authors improves the performance and efficiency of fuzzy-rule based routing algorithms using the modified clonal selection algorithm (CLONALG-M). The modified version of the clonal selection algorithm is applied to find the nearest form of the output membership functions, which help to improve the overall fuzzy routing algorithm performance. The authors [41] proposed an Energy-efficient fuzzy logic cluster head (EEFL-CH) algorithm, which is the improvement of the LEACH protocol. The aim of this algorithm is to reduce energy consumption while increasing the network lifetime by using fuzzy logic. CH selection is based on three fuzzy parameters residual energy, expected efficiency, and closeness to the base station. The author LEE [42] proposed a fuzzy-logic-based clustering approach with an extension to the energy predication, which enhances the network lifetime by evenly distributing the workload. This approach uses two fuzzy parameters, residual energy, and expected residual energy, for elected CHs.

## 2.3 System Model

In this chapter we consist the network model along with the energy model and terminologies used in it.

### 2.3.1. Network Model

In our network model, we have considered a network where a number of homogeneous sensor nodes are randomly deployed in an area and BS is located outside of the network area. For our proposed network model, we make some assumptions, which are as follows.
1) All the sensor nodes and the base station are considered to be static after deployment.
2) Sensor nodes can join just a single CH inside its imparting range.
3) The BS has no energy limitation.
4) Wireless connection is symmetric and bidirectional.

5) Initially, all the sensor node contains the same amount of energy.

## 2.3.2. Energy Model

Consider nodes are distributed uniformly in M*M region. If there is a cluster k, then the average is N/K nodes per cluster. The area engaged by each cluster is approximately $M^2/K$. The required squared distance from the nodes to the CH is given by

$$E\left[ d_{toCH}^2 \right] = \rho(x, y) dxdy \tag{2.1}$$

We assume that this area is a circle with a radius $R = \left(\dfrac{M}{\sqrt{\pi K}}\right)$ and $\rho(r,\theta)$ is constant for r and $\theta$, so equation 1 simplifies to:

$$E\left[ d_{toCH}^2 \right] = \rho \int_{\theta=0}^{2\pi} \int_{r=0}^{\frac{M}{\sqrt{\pi K}}} r^3 \partial r \partial \theta = \frac{\rho}{2\pi} \frac{M^4}{K^2} \tag{2.2}$$

If the density of the nodes is uniform throughout the cluster area, then the diameter

$$D = \frac{2M}{\sqrt{\pi K}}$$

and
$$E\left[ d_{toCH}^2 \right] = \frac{1}{2\pi} \frac{M^2}{K}$$

The Average distance between CH nodes to BS:

$$D_{toBS} = \int_A \sqrt{x^2 + y^2} \frac{1}{A} \partial A = 0.765 \frac{M}{2} \tag{2.3}$$

$$D_{toBS} = \frac{0.755M}{2} \tag{2.4}$$

Now first, we consider all cluster are equal in size, so the radius is R, and the area of each cluster is $\dfrac{M}{K}$, therefore the radius of a cluster is $R = \dfrac{M}{\sqrt{\pi K}}$, Now the diameter of the cluster or distance between two cluster head is

$$D = \frac{2M}{\sqrt{\pi K}} \tag{2.5}$$

The energy model considered in our work is alluded from [39] as shown in Fig. 2.1. The signal-to-noise ratio (SNR) during broadcasting l-bit packet over distance d so the energy dissipated by radio is determined as:

$$E_{TX}(l,d) = \begin{cases} lE_{elec} + l \in_{fs} d^2, & if \ d \leq d_o \\ lE_{elec} + l \in_{amp} d^4, & if \ d \geq d_o \end{cases} \quad (2.6)$$

where, $E_{elec}$ denotes energy depleted per bit to operate the electronic circuitry, $\in_{fs}$ and $\in_{amp}$ shows the energy consumption by free space and multi-fading channel, d indicate the distance among sender and receiver and $d_0$ denotes the threshold transmission distance.

The threshold distance $d_0$ is given by equating the equation for $d = d_o$

$$lE_{elec} + l \in_{fs} d_o^2 = lE_{elec} + l \in_{amp} d_o^4 \quad (2.7)$$

$$d_o = \sqrt{\frac{\in_{fs}}{\in_{amp}}} \quad (2.8)$$

To obtain l-bit message, the radio uses:

$$E_{RX}(l,d) = lE_{elec} \quad (2.9)$$

So the total energy consumption we get:

$$E_t = l\{2 N_s E_{elec} + N_s \in_{fs} d_{toCH}^2 + N_s E_{DA} + k \in_{fs} D^2 + \in_{mp} k \, d_{toBS}^4\} \quad (2.10)$$

To find the optimal number of cluster, we differentiate equation (9) with k and put the value 0.

$$K_{opt} = \frac{M}{d_{toBS}^2} \sqrt{\frac{N_s}{2\pi}} \sqrt{\frac{\in_{fs}}{\in_{mp}}} \quad (2.11)$$

where M is the side of the deployment area, so area A= M*M $m^2$ .

$E_{elec}$ indicates energy consumed in transmitting or receiving one bit, depends upon coding and modulation.

$\in_{fs}$ indicates free space loss coefficient.

$N_s$ shows sensing nodes and

$E_{DA}$ is aggregation energy.

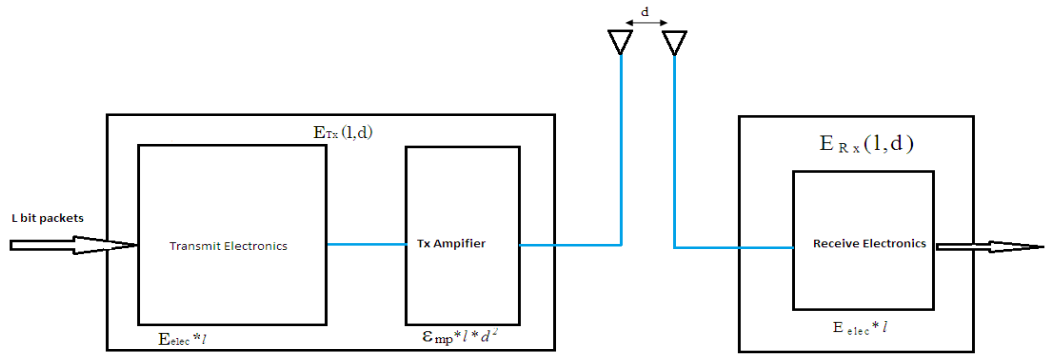

Fig. 2.1 Radio model

## 2.4 Unequal Clustering

In the wireless network, to reduce energy consumption, sensor nodes chose a CH to forward their data as shown in Fig. 2.2. As the distance from the BS is one of the main factors for a node to be selected as a CH, a number of CM nodes can choose them for their data forwarding leading the CH to drain its battery rapidly. As the node's distance from BS increases, the energy required in data transmission also increases, leading to the node's rapid battery drainage. To reduce CH's rapid battery drainage, we propose an unequal clustering scheme where the radius of a CH varies depending on their distance from BS. i.e., if the cluster is near the base station, then the cluster size will be small, and if the cluster is far from the BS, then the cluster size will be large. In this scheme, we first select candidate CHs, and later, based on other properties, we select final CHs for data transmission. The scheme is discussed as follows.

In this scheme, based on a probabilistic approach, we select initial CH and provisional CH. For initial CH selection, in each clustering round, every sensor node creates an arbitrary number between the two binary numbers (0 and 1). If the arbitrary numbering for a specific node is lower than the threshold (T), this node spontaneously becomes an initial CH. The competition radius from every provisional CH dynamically changes in the proposed model. This is because the proposed model utilizes the node concentration, distance to the BS, and residual energy in calculating the relevant competition radius. Therefore, it is logical to reduce the network service area for a CH whose residual energy is proportionally decreasing. The sensor node rapidly runs out of its battery if the competition radius does not reduce while its residual energy reduces. Radius calculation is achieved by utilizing preset fuzzy IF-THEN delineated regulations to manage the uncertainty. To assess the regulations, the Mamdani Method (one of the most utilized methods) is utilized as a fuzzy inference technique [10].

The Center of the area (COA) practice is used for the de-fuzzification process in competition radii. In each round, initial CHs are selected by creating a random number that is allocated to every node. If the produced random number allocation is less than the threshold value (TH) of adjacent nodes allocated in Eq. 12, then that node becomes a provisional CH.

$$TH = P/(1 - P*(r \bmod 1/p)) \tag{2.12}$$

where r is the value of the present round of selection, P is the preferred percentage of CH (e.g., P = 0.05). This section will describe the unequal clustering process by making use of fuzzy logic. Different from other studies, three linguistic variables are used. These linguistic variables used are residual energy in the provisional CH, distance to the BS, and concentration.

• **Input Distance to BS-** Distance to the BS is a crucial metric in avoiding energy holes. The shorter distance of a CH to the BS has a smaller number of cluster member nodes, hence spends considerable energy to relay data packets. However, distant CHs can have large clusters because data is propagated in a multi-hop communication process.

• **Residual Energy-** CH node functions more than its member nodes. These functions include data collection, aggregation, and transmission to the BS. It should, therefore, have enough energy to support all relevant functions.

• **Concentration-** The number of adjacent nodes in the vicinity of the preferred CH.

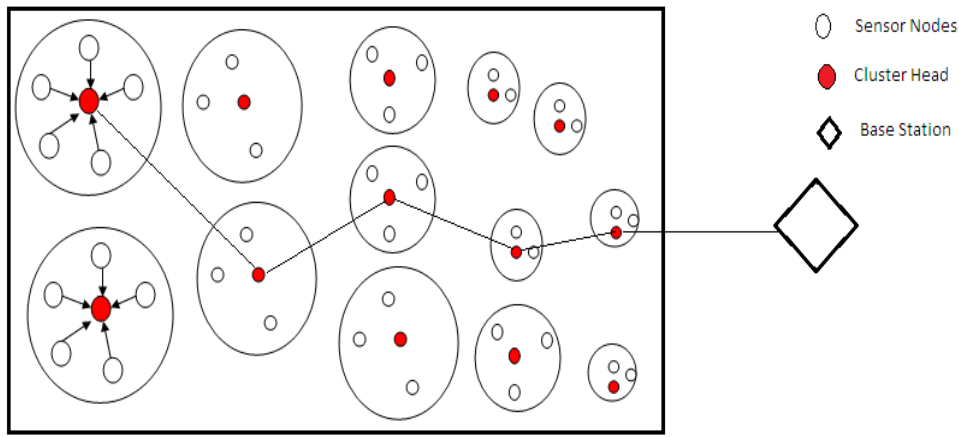

Fig. 2.2 Proposed model

This section includes a description of the unequal clustering process using fuzzy logic. In this paper, three variables were used. The variables used include residual energy of the provisional CHs, distance to the BS, and concentration of nodes adjacent to the CH. The existing fuzzy-based unequal clustering process has one output variable only. The suggested process includes two output variables. These are CH probability and its competition radius. The latter variable is made up of eight linguistic values: small, very small, rather small, medium, medium-small, large, rather large, medium-large, and very large. The values with trapezoidal membership functions are very large and very small. The rest have triangular functions. The former variable is made up of seven fuzzy linguistic values. They include poor, very poor, average, below average, above average, strong, and very strong. The values with trapezoidal membership functions are very strong and very poor. Triangular functions characterize the rest. The careful selection of membership function extent is supposed to subject all the nodes to a repeated experimental analysis on different network magnitudes. The fuzzy theory engine fuzzifies the crisp data values into suitable linguistic variables. The engine makes use of availed membership functions. The Mamdani practice is utilized to come up with regulations with which the fuzzified data values are refined. Twenty-seven regulations are eventually specified (as in Table 2.1), which blend various linguistic variables. The output variable is also a fuzzy value. The trapezoidal and triangular membership function used in our FIS is shown in the following equations:

$$\mu_{A1}(x) = \begin{cases} 0 & x \leq a_1 \\ \dfrac{x-a_1}{b_1-a_1} & a_1 \leq x \leq b_1 \\ \dfrac{c_1-x}{c_1-b_1} & b_1 \leq x \leq c_1 \\ 0 & c_1 \leq x \end{cases} \tag{2.13}$$

$$\mu_{A2}(x) = \begin{cases} 0 & x \leq a_2 \\ \dfrac{x-a_2}{b_2-a_2} & a_2 \leq x \leq b_2 \\ 1 & b_2 \leq x \leq c_2 \\ \dfrac{d_2-x}{d_2-c_2} & c_2 \leq x \leq d_2 \\ 0 & d_2 \leq x \end{cases} \tag{2.14}$$

The COA method is utilized to de-fuzzify the output variable into a crisp variable. After completing the CH selection, non-CH nodes unite with the nearest CH. The COA method is used in the given equation

$$COA = \dfrac{\int \mu_A(x) x\, dx}{\int \mu_A(x)\, dx} \tag{2.15}$$

Table 2.1 Fuzzy Rules

| Distance | Residual Energy | Concentration | Cluster Radius | CH Choice |
| --- | --- | --- | --- | --- |
| Close | Less | High | Very small | Very poor |
| Close | Less | Med | Small | Poor |
| Close | Less | Low | Rather small | Below average |
| Close | Avg | High | Small | Avg |
| Close | Avg | Med | Rather small | Below avg |
| Close | Avg | Low | Medium small | Poor |
| Close | High | High | Rather small | Very strong |
| Close | High | Med | Small | Strong |
| Close | High | Low | Medium small | Above avg |
| Far | Less | High | Medium small | Avg |
| Far | Less | Med | Rather small | Below avg |
| Far | Less | Low | Small | Poor |
| Far | Avg | High | Medium large | Below avg |
| Far | Avg | Med | Medium | Avg |
| Far | Avg | Low | Medium small | Below avg |
| Far | High | High | Medium large | Strong |
| Far | High | Med | Medium | Above avg |
| Far | High | Low | Medium small | Avg |
| Farthest | Less | High | Large | Poor |
| Farthest | Less | Med | Medium large | Very poor |
| Farthest | Less | Low | Medium | Below avg |
| Farthest | Avg | High | Rather large | Avg |
| Farthest | Avg | Med | Large | Below avg |
| Farthest | Avg | Low | Medium large | Above avg |
| Farthest | High | High | Large | Very strong |
| Farthest | High | Med | Rather large | Strong |
| Farthest | High | Low | Very large | Above avg |

## 2.5 Cluster Formation

Algorithm 1 gives the pseudocode for CH selection and cluster formation in each round. The algorithm is explained as follows. At the beginning of every round, each sensor node creates an arbitrary value between 1 and 0. If the arbitrary value created by a specific node is less than the threshold (TH) value, that node becomes a provisional CH. This algorithm utilizes the node's distance to the BS, residual energy, and concentration to determine the relevant competition radius. It should be noted that in some cases, the hotspot problem might occur with the increasing number of clusters and inter cluster routing. With this problem, a number of clusters may be generated with too small radius.

As a solution to this problem, we set a lower radius to prevent the network from becoming divided into too many clusters. The provisional CH will then compute radius and chance. Every node relays a discovery CH-MSG in the cluster radius to generate its routing table carrying a list of adjacent nodes and their residual energy. The CH will broadcast CH-MSG to all adjacent nodes in its radius, which is determined by FIS. This provisional CH-MSG will have information about the node's chance value and id. The Provisional CH with the highest chance value within the cluster is declared the Prime cluster head (PCH). The PCH then relays ELECTED-CH-MSG to adjacent nodes. The rest of the nodes that do not become the PCH will relay JOIN-CH-MSG to the adjacent CH.

Finally, the PCH transmits a message which contains a time slot table to its member nodes. According to this slot table, the member nodes send raw data to the PCH. After the completion of clustering, the packet transmission is similar to EAMMH and EAUCF schemes. Thus, the data transmission and synchronization process in the cluster is not introduced in detail in this work.

There is a dynamic change in the provisional CH competition radius of this suggested algorithm, as it utilizes energy, shorter distance to the BS, and low concentration in the parameters for CH chance calculation. Reducing the CH radius is the best option to reduce its energy consumption. The preset fuzzy IF-THEN regulations determine the CH chance. By using the fuzzy process, unpredictability is integrated into the WSN and is computed efficiently.

The input variable takes the fuzzy values that are equivalent. In the initial input variable, the node's distance to the BS is characterized by close, far, and farthest as the fuzzy linguistic values. Farthest and close have trapezoidal functions, while far has triangular functions.

**Algorithm 1 CH Selection and Clustering**

1: **TH** ← Threshold probability for selection of a provisional CH
2: **for** each node **do**
3:     Provisional CH = False
4:     $\mu$ ← rand (0,1)
5:     **if** ( $\mu$ < TH) **then**
6:         Provisional CH = True
7:     **else** Provisional CH = False
8:     **end if**
9:     **if** Provisional CH=True **then**
10:         **for** each provisional CH **do**
11:             Using if then mapping rules, calculate radius and CH chance
12:             Radius= Fuzzy logic (distance, residual energy, node concentration)
13:             CH chance = Fuzzy logic (distance, residual energy, node concentration)
14:         **end for**
15:         Inside the range of Provisional CH send CH message (ID, chance) to other CHs
16:         Select Provisional CH having maximum CH chance as Prime CH
17:     **end if**
18:     **if** Provisional CH 6= Prime CH then
19:         Provisional CH =node
20:     **end if**
21:     Send Prime CH message (ID, chance)
22:     **for** (each node) **do**
23:         join with Prime CH having maximum as a cluster member
24:     **end for**
25:     **for** each Prime CH **do**
26:         calculate the distance among all clusters
27:         **if** (distance between Prime CH and BS > min (Distance between Prime CH - Prime CH')) **then**
28:             transmit to BS
29:         **else** (transmit to Prime CH)
30:         **end if**
31:     **end for**
32: **end for**

The next input variable, node's residual energy, is characterized by less, average, and high as the linguistic values. High and less have trapezoidal functions. The average has a triangular membership function. The third input variable is concentration, which is made up of low, medium, and high as the linguistic values. Low and high have trapezoidal functions. Medium has triangular functions. The non-CH nodes relay sensed data to their specific CHs after clusters are formed. The eventual CHs then combine all the received input and relay it farther to another CH using multi-hop transmission, and then CH sends the data to the sink.

## 2.6 Simulations and Results (Scenario 1)

The proposed algorithm has been assessed using MATLAB because its Fuzzy Toolbox examines all fuzzy membership functions, hence suitable for use. This suggested algorithm was simulated using MATLAB. One hundred sensor nodes were reviewed after distribution over a $100x100m^2$ area. Assumptions made included initial energy in every node as 0.5J. The MATLAB simulation frameworks utilized in the suggested system are tabulated in Table 2.2. Shown in Table 2.3 is the suggested protocol outperforming in FND. The network duration has been evaluated as the period at the beginning of the operation to the initial node death or the eventual node death.

Table 2.2 Simulation Parameters

| Parameter | Value |
|---|---|
| Area | 100*100 |
| Nodes | 100 |
| $\varepsilon_{mp}$ | 0.0013 pj/bit/m4 |
| $\varepsilon_{fs}$ | 10 pJ/bit/m2 |
| $E_{elec}$ | 50 nJ/bit |
| Packet size | 4000 bits |
| Initial Energy | 0.5j |
| Control Packet Size | 200 bits |

We tested the suggested algorithm extensively. Experimental results were presented in a tabulated design. The suggested algorithm is compared to EAUCF, LEACH, TTDFP Tier-1, and EAMMH algorithms. The results show the suggested algorithm performs greater than EAUCF, LEACH, EAMMH, TTDFP Tier-1 algorithms in both scenarios. The limited energy is the major constraint in the WSN. If energy-consumption between the nodes varies, some nodes will finish their energy earlier than others, hence making the network unstable.

Fig. 2.3 shows the measured stability of the WSN. The measurement is achieved by comparing the average energy consumed in the network in each round. Therefore, the suggested algorithm consumes less energy compared to the older algorithms, such as EAMMH, EAUCF, LEACH, and TTDFP Tier-1 algorithms.

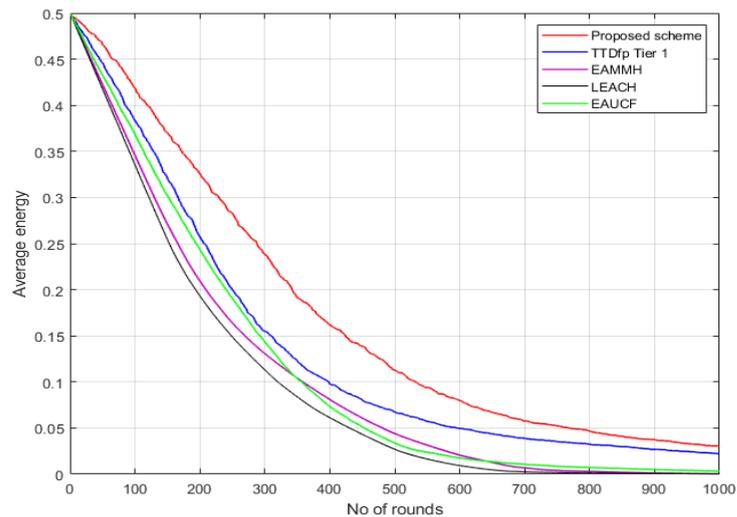

Fig 2.3 No. of rounds vs average energy

Several calculations have previously been utilized in other literature to explain the network duration of WSNs. The Half Node Die (HND) and First Node Die (FND) metrics are some of the commonly utilized calculations in EAUCF [13]. HND and FND include many rounds. The genesis of the network operation proceeds with the initial node until it runs out of battery. The other half of the rest of the nodes also run out of battery, respectively. Fig. 2.4 and fig. 2.5 demonstrate the simulation results of the network duration (HND and FND). Therefore, the suggested algorithm will perform better than previously considered algorithms (EAMMH, LEACH, TTDFP Tier-1, and EAUCF) for HND and FND calculations. Among the previously utilized algorithms, LEACH has the lowest performance. This is because LEACH utilizes a probabilistic method to choose CHs. For 500 rounds of cluster functions, nodes in LEACH start to run out of battery at 118 rounds. On the other hand, nodes in the EAMMH protocol start to run out of battery at 164 rounds, while the first node in the EAUCF protocol starts to run out of battery at 109 rounds, and the TTDFP Tier-1 protocol starts to run out of battery at 285 rounds. However, our suggested algorithm includes the first node running out of battery at 353 rounds.

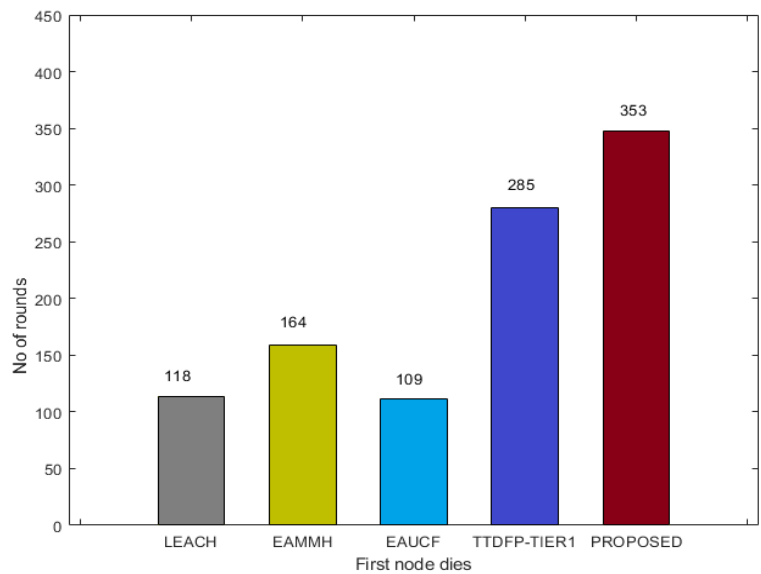

Fig 2.4 FND vs no. of rounds

The FND suggested algorithm has more outstanding performance than the TTDFP Tier-1 algorithm by 123.85%, EAUCF algorithm by 323.85, LEACH algorithm by 299.15%, and EAMMH algorithm by 215.24%. In fig. 2.5, half node died at 372 in LEACH, In EAMMH, half node died at 416, in EAUCF, half of the node died at 515, in TTDFP Tier-1 HND at 593, and in the proposed algorithm, half node died at 610, which is better than comparing to all these algorithms. The HND suggested algorithm has greater performance than the TTDFP Tier-1 by 102.86%, EAUCF algorithm by 118.44%, LEACH algorithm by 163.97%, and EAMMH algorithm by 146.63%.

Fig. 2.5 presents the number of sensor nodes that ran out of battery versus their respective number of rounds. In the suggested algorithm, the nodes that have run out of battery are less. The proposed algorithm considers all the possibilities in the selection of CHs. Therefore, energy usage by the WSNs is less, resulting in the recorded lower number of nodes running out of battery.

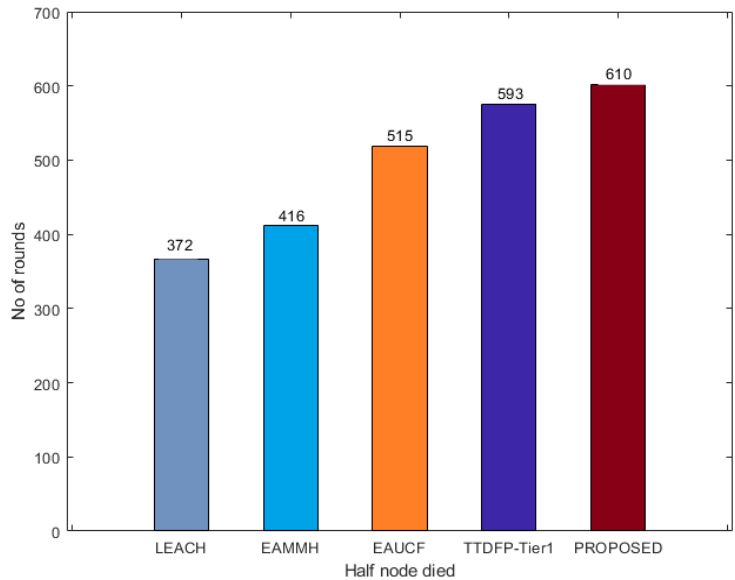

Fig 2.5 HND vs no. of rounds

Fig. 2.6 and fig. 2.7 expresses that the first node died, and the Last Node died in the different network areas vs. the round number. The proposed algorithm performance is better than LEACH, EAMMH, EAUCF, and TTDFP Tier-1. The purpose of taking different network sizes is to establish the effect of the density of the node on the suggested algorithm. Equal clustering is used to determine a low-energy adaptive hierarchy. Therefore, balancing of workload between adjacent nodes is not performed in this suggested algorithm. EAMMH (which is an energy-aware unequal clustering fuzzy protocol) balances the workload. However, its CH choice is determined by one parameter. This makes energy consumption to be high. The suggested protocol balances energy usage across the network. Forming unequal clusters achieves this. Besides, CH selection is achieved using the fuzzy method. In conclusion, the suggested algorithm performs better in all metrics than the previously used algorithms.

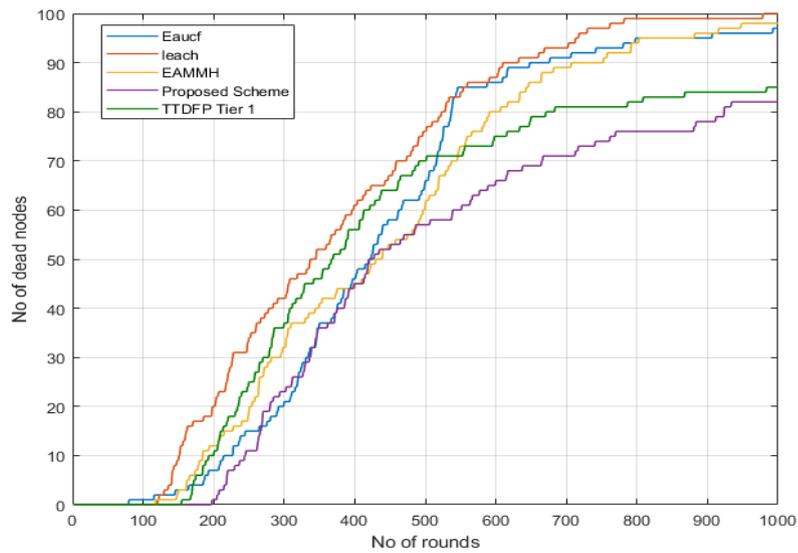

Fig 2.6 Dead nodes vs no of rounds

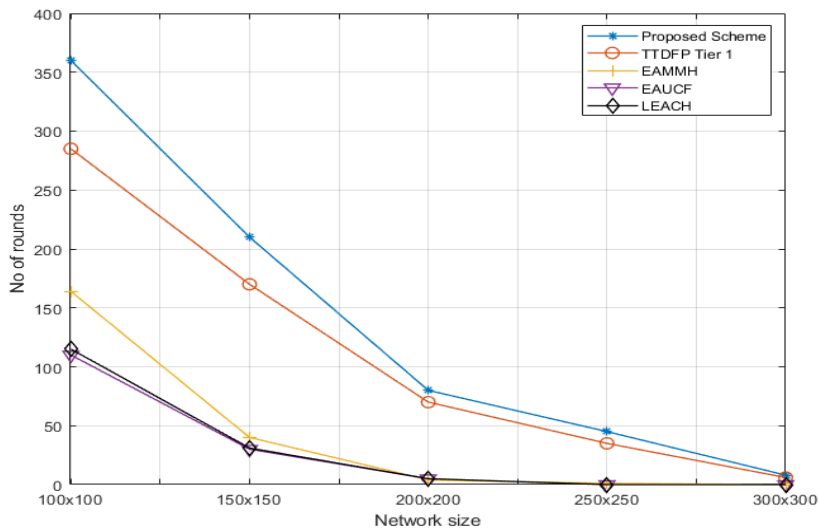

Fig 2.7 no of rounds vs network size

## 2.7. Energy dissipation

The following table 2.3 shows the energy dissipation of the proposed protocol with other algorithms at different network sizes. It can be seen that the energy dissipation of the proposed protocol is less with respect to other protocols as LEACH, EAMMH, and EAUCF, which leads to prolonging network lifetime. Also, the smaller cluster size near the sink in the proposed protocol balances the load among the nodes as well as solves the energy hole problem.

Table 2.3 Energy dissipation

| Network size | Algorithm | FND energy dissipation | HND energy dissipation |
|---|---|---|---|
| 100*100 | LEACH | 0.3032 | 0.0871 |
|  | EAMMH | 0.2983 | 0.0738 |
|  | EAUCF | 0.3587 | 0.0422 |
|  | TTDFP | 0.2573 | 0.0671 |
|  | PROPOSED | 0.2345 | 0.0551 |
| 150*150 | LEACH | 0.3515 | 0.1362 |
|  | EAMMH | 0.3576 | 0.1222 |
|  | EAUCF | 0.4077 | 0.1187 |
|  | TTDFP | 0.2963 | 0.0852 |
|  | PROPOSED | 0.2548 | 0.0637 |
| 200*200 | LEACH | 0.3683 | 0.1237 |
|  | EAMMH | 0.3934 | 0.1333 |
|  | EAUCF | 0.4758 | 0.1371 |
|  | TTDFP | 0.3581 | 0.1394 |
|  | PROPOSED | 0.3202 | 0.0760 |
| 250*250 | LEACH | 0.4189 | 0.1246 |
|  | EAMMH | 0.4380 | 0.1416 |
|  | EAUCF | 0.4996 | 0.1424 |
|  | TTDFP | 0.4037 | 0.1269 |
|  | PROPOSED | 0.3733 | 0.1013 |
| 300*300 | LEACH | 0.3481 | 0.1468 |
|  | EAMMH | 0.4362 | 0.1501 |
|  | EAUCF | 0.4698 | 0.1584 |
|  | TTDFP | 0.3911 | 0.1395 |
|  | PROPOSED | 0.3830 | 0.1006 |

## 2.8. Scenario 2

In this scenario, we try to assess the performance of clustering independently. The distribution of a total of 1000 sensor nodes and BS in a 1000m*1000m monitoring area. We assume the initial energy is 0.5J. The simulation frameworks utilized in the proposed system are shown in table 2.4. In this scenario, fig. 2.8 shows the average energy of the nodes when the network executes 50, 100, and 150 rounds, which reflect the energy consumption of the network. In the graph, the proposed algorithm is always higher than the other algorithms and performs approximately 99.99% better than LEACH, nearly 83.4% better than EAMMH, 46.37% better than EAUCF, and 19.5% better than TTDFP Tier-1. This study shows that the proposed algorithm has lower energy consumption that means its network lifetime is better and stable than other these comparing algorithms.

Table 2.4 Simulation Parameters of scenario 2

| Parameter | Value |
|---|---|
| Area | 1000m*1000m |
| Nodes | 1000 |
| Packet size | 4000 bits |
| $\varepsilon_{mp}$ | 0.0013 pj/bit/m4 |
| $\varepsilon_{fs}$ | 10 pJ/bit/m2 |
| $E_{elec}$ | 50 nJ/bit |
| Initial energy | 0.5j |
| $E_{DA}$ | 5nj/bit/signal |
| Control packet size | 200 bits |

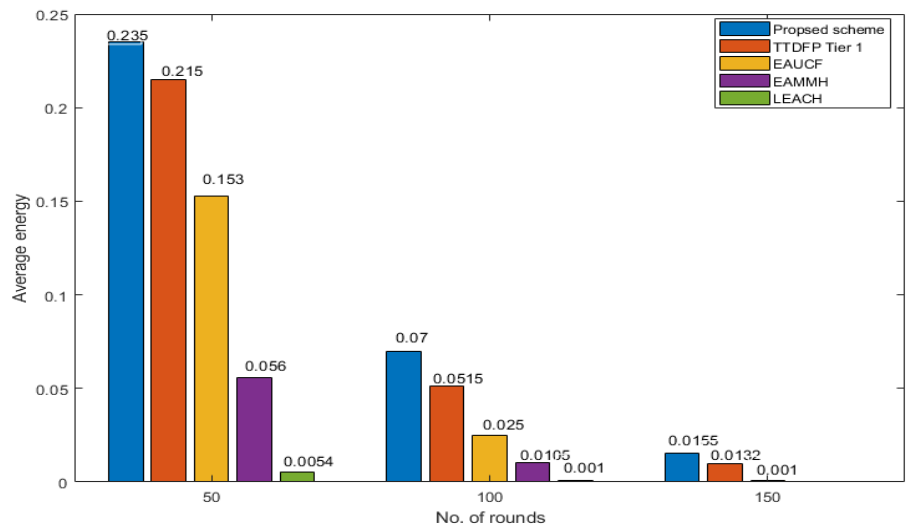

Fig 2.8 Average energy vs no. of rounds

Fig. 2.9 shows the number of alive nodes in this scenario, and the proposed algorithm performance is better than comparing algorithms. Considering 300 rounds in this result and get LEACH algorithm nodes died quickly in the large scale WSN because LEACH sends data directly from CHs to BS. Other EAUCF, EAMMH, and TTDFP Tier-1 algorithm nodes also low survival time.

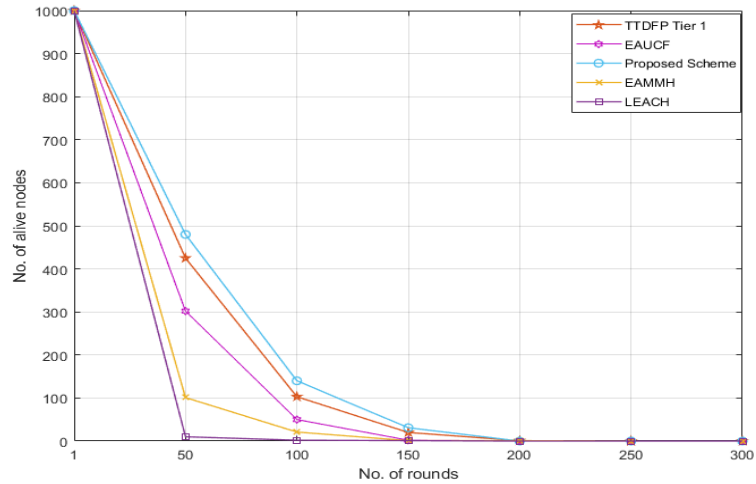

Fig 2.9 No. of alive nodes vs no of rounds

Fig. 3.0 shows the number of rounds vs. half node died in scenario 2. The proposed algorithm has better performance than other comparing algorithms. The proposed algorithm performs 204.34% better than LEACH, 188% better than EAMMH, 146.87% better than EAUCF, and 117.5% better than TTDFP Tier-1.

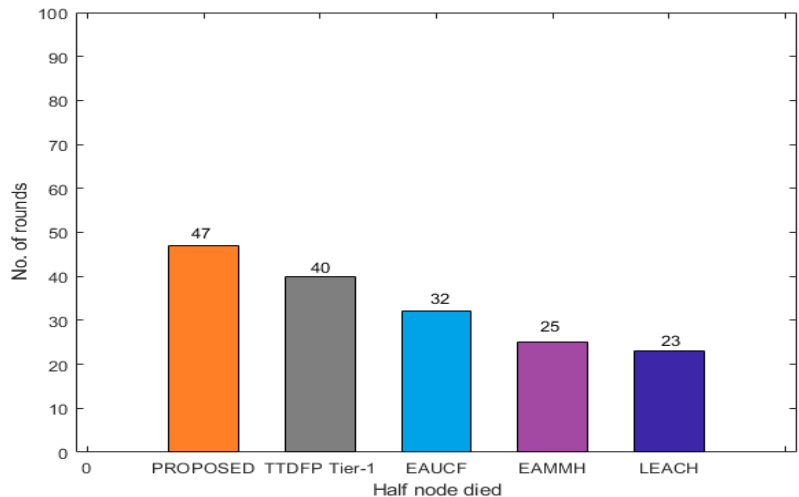

Fig 2.10 No. of rounds vs half node died

## 2.9 Chapter Conclusion

We proposed a fuzzy logic-based protocol because it performs better than other comparing algorithms. Some of the assumptions made include equally significant stationary networks of the existing nodes. Regarding this assumption, we introduced the new algorithm in three subsequent steps. The initial step is to apply multi-hop communication to reduce energy consumption. The step is followed by an offline phase, where an unequal clustering algorithm solves the energy hole issue by reducing the sizes of clusters near the BS and determining the competition radius through the fuzzy logic protocols.

　The final step utilizes fuzzy logic. After the distance to a BS, node residual energy and concentration have been used to select CHs. This final step helps in the distribution of workload among the CHs. Simulation results demonstrated that our process is useful for implementations that need minimization of energy consumption, load balancing, and prolonged network duration. Overall, the proposed algorithm is a better protocol compared to TTDFP Tier-1, EAMMH, EAUCF, and LEACH clustering algorithms. In our future work, we will use the interval type-2 fuzzy logic theory to solve the secure clustering protocol for mobile ad-hoc networks and the light weight secure mechanism will be our research focus.

# Chapter 3 Energy-Efficient Cluster Head Election for WSNs

## 3.1 Chapter Overview

Routing is one of the primary technologies in wireless sensor networks (WSNs). Many studies have shown that cluster-based WSN routing protocols excel in network topology control, energy minimization, load balancing, and so on. Recently, there has been a considerable amount of research on clustering protocols for WSN, and noticeable progress has been made in this field. As the WSN consist a large number of randomly deployed nodes with data storage and data transmitting capabilities, clustering routing schemes provide an efficient way to solve energy consumption and topology management problems. The clustering schemes also provide the ease of operation, extend device life and lower the latency of data transfer [3].

In WSN, compared to sensing and encoding, transmitting information requires more energy that makes the energy efficiency a challenging and sever issue in the network. To reduce the energy usage, in clustering scheme the network nodes are clustered together based on specific parameters to form a cluster. In some clusters, a node is selected as cluster head (CH) among the cluster members (CM). The CH gathers the information form the CMs and forwards it to BS via single-hop or multi-hop links. This cluster-based configuration is one of the ways to cope with the energy savings of sensor node systems [43].

Considering the size of the clusters, there are mainly two kinds of clustering schemes, equal clustering scheme and unequal clustering scheme [44]. In equal clustering scheme, each formed cluster has equal radius. The drawback with equal clustering scheme is that the CHs near the BS are selected by a heavy number of devices to forward their data which lead to a premature energy depletion of such CH. To overcome this problem the unequal clustering scheme is utilized. In unequal clustering scheme, the size of each cluster is decides based on some parameters. In general, to reduce the data traffic burden, the clusters near the BS are made narrower than clusters away from the BS.

In this chapter, stressing the scalability of the network, and energy consumption of sensor nodes, we propose a CH selection algorithm based on type-2 fuzzy logic that also enhances the network life time. In this algorithm we choose the CHs based on type-2 fuzzy logic taking its distance from BS and residual energy as two inputs of fuzzy logic. To save the CHs from quick energy drainage, we also design a unequal clustering scheme multi hop routing techniques to respectively choose the size of clusters and forward the data to the BS.

## 3.2 Related Works

In recent years, various kind of routing protocols for WSN has been designed in many research works. In this section we divide and summarize most of these protocols into clustering-based routing protocols and fuzzy based routing protocols.

### 3.2.1 Clustering Based Routing Protocols

The low-energy adaptive clustering hierarchy (LEACH) protocol in [45] is a well-known standard protocol that elects the cluster heads based on a probabilistic model. In this protocol, the sensor nodes are elected as CH based on their remaining energy. This protocol improves the life of the networks but in some situations in this protocol sensor nodes with lower energy are selected as CH. The energy of these CHs depletes quickly that, creates energy hole problem in the network. The authors in [21] propose a hybrid energy efficient distributed (HEED) protocol similar to LEACH, in this protocol, the CHs are also selected based on a probabilistic model. However, this protocol uses two clustering parameters called as primary and secondary. The primary parameter selects CHs based on the residual energy of the nodes while the secondary parameter selects CHs based on the node degree or node proximity. In [46], the authors propose a energy-aware multi-hop multi-path hierarchical (EAMMH) routing protocol. For developing this protocol, the authors utilize the features of energy aware routing and multi-hop intra-cluster routing together.

### 3.2.2. Fuzzy Based Protocols

In [38], Bagci and Yazici propose an energy aware unequal clustering fuzzy (EAUCF) scheme. In this scheme, the authors use a rule-based fuzzy logic to select the CHs with the help of two linguistic variables named as the distance to BS, and residual energy. The downside of this algorithm is that it might select some CH nodes which have less energy and doesn't consider any parameter to cope with inter-communication cost as well.

   Stressing the drawback of EAUCF, The authors in [39] present a multi-objective fuzzy clustering algorithm (MOFCA). In this algorithm along with distance to BS and residual energy variables, the author includes one more variable as node density to select the cluster head chance and competition radius. In fuzzy logic based unequal clustering (FBUC) scheme [47], the author considers a new parameter as node's degree along with the residual energy and distance to the BS to measure CH chance and compute the radius. The benefit of this protocol is that it increases the life of the network but conserves more energy in the intra-cluster and data transmission process. The authors A. Lipare and D.R.Edla in [48] propose an algorithm for the election and construction of CHs where they use fuzzy logic with three input parameters i.e., distance to BS, node degree, and remaining energy to select CHs. After selection of CHs, the sensor nodes are distributed to their separate CHs with the closest separation and the accessible size of CH.

   In [38], the authors present a novel two-tier distributed fuzzy logic-based (TTDFP) protocol which also enhance the lifetime of the network. It is a distributive protocol that drives and compares wireless sensor network applications efficiently. To optimize the performance of sensor networks, the author presents an optimization method to adjust the parameters used in the fuzzy logic. In [42], propose a fuzzy logic-based clustering approach with an expansion of energy forecast. This protocol prolongs the lifetime of network by equitably circulating the remaining workload. For clustering scheme, the authors uses fuzzy logic with two parameters, residual energy and required residual energy and compute a selection chance for each CH. In [49], the authors use a non-probabilistic approach to select the CHs by considering the residual energy of the nodes. The fuzzy input parameters distance to BS and node density are used to compute the radius of each cluster.

### 3.3. System Model

In this section we explain our proposed network model.

### 3.3.1. Network Model

In this chapter, we consider a WSN with a BS at centre of the cell and a number of sensor nodes deployed around the BS randomly. For data transmission we consider that some nodes may elected as the CH and rest of the nodes can transmit their data to the BS thorough the suitable CHs. In this network each sensor node sends the data to the nearest CH, each CH gathers data from its cluster members, coagulates, and sends compressed data to the BS via other CHs along with their own data using multi-hop transmission.

The considered assumptions of this network model are as follows:

1. All the nodes are homogeneous and deployed randomly in the WSN.
2. All nodes are stationary after deployment and have the same initial energy.
3. After installation, the base station is stationary and inside the network.

In our proposed scheme, the radio model is similar to [14] that is considered to predict the node's energy consumption. This model grants free space and multipath fading channels based on the distance between the deliver node and the receiving node.

Let us consider that for transmitting the $l$ bit message at $d$ distance, $E$ energy dissipation is required, then the function $E(l,d)$ can be represented as

$$\mathrm{E}(l,d) = E_{TX-elec}(l) + E_{TX-amp}(l,d) \tag{3.1}$$

where $E_{TX-elec}$ represents the energy dissipated by transmitter to run the radio electronics and $E_{TX-amp}$ represent the energy dissipated to run the amplifier at the transmitter.

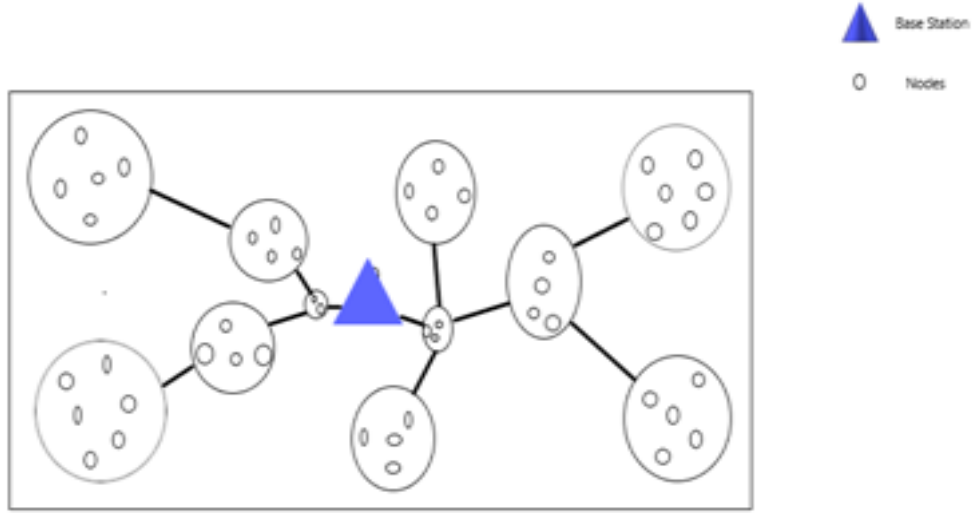

Fig 3.1 proposed network model

The amount of total energy consumption required can further be extended as.

$$E(l,d) = \begin{cases} lE_{elec} + l\epsilon_{fs}d^2, & if\ d \leq d_o \\ lE_{elec} + l\epsilon_{amp}d^4, & if\ d \geq d_o \end{cases} \quad (3.2)$$

where $E_{elec}$ represents the energy consumption by the circuit, $\varepsilon_{fs}$ denote the free space energy consumption, $\varepsilon_{mp}$ is denotes multipath fading channel and $d_0$ represent the threshold distance between transmitter and receiver.

### 3.3.2. Interval type-2 Fuzzy Logic

In comparison to type-1 fuzzy logic, type-2 fuzzy logic is more comfortable that handles the uncertainty more effectively. The interval type-2 fuzzy logic can better handle the sensor network's uncertainty than the traditional type-1 fuzzy sets. This interval type-2 FIS simplifies the calculation and is more appropriate for the sensor network. In the proposed scheme, we consider two factors, the distance to the base station and the residual energy as the input of type-2 fuzzy logic that affect the CH election and competition radius.

### 3.3.3. Fuzzifiers

1. Distance to BS

The distance to BS is an important input for type-2 fuzzy logic to select CH chance and determine the competition radius of CHs. Since for data transmission each node sends the data to the BS through the CH, the pretended CH near to the BS have high probability to be selected as the final CH of a large number of nodes than the pretended CH that are far from the BS. In this situation, the energy of the CHs near to the BS drains quickly. To cope this situation, we design our model in such a way that the CHs near to the BS have smaller radius and the CHs far from BS have larger radius.

For selection of the CH radius, we divide the distance to the base station into three membership functions, proximate, moderate, and far. At the initial stage of the network, the transmitted signal is received by each node from the BS, and the distance to the base station is calculated by the value of the signal strength (RSSI). Here, from the following equation (3.3), we can calculate the distance between node $S_j$ and BS.

$$DB = \frac{d(S_j, BS)}{d_m} \quad (3.3)$$

where, $d_m$ indicates maximum communication distance of the network, and $d(S_j, BS)$ shows the distance between BS and sensing node.

2. Residual Energy

For the fuzzy logic input, the residual energy is classified into three membership function low, medium, and high. It is crucial to consider the residual energy of the nodes while choosing the CH chance and radius. The given equation shows the calculation of the residual energy where $E_P$ is the current energy value of each node, and $E_o$ indicates the initial energy:

$$\mathrm{Res}.Energy = \frac{E_P}{E_0} \quad (3.4)$$

### 3.3.4. Rule Base

The type-2 interval based fuzzy logic has two output variables, the node competition radius and the CH chance. To minimize computational complexity, the resultant rule is divided into six membership functions, which are very weak (VW), weak (W), medium (M), higher medium (HM), strong (S), very strong (VS). The rule base including nine fuzzy rules, are shown in Table 3.1.

Table 3.1 TYPE-2 Fuzzy rules

| Distance  | Residual Energy | Radii      | Chance |
|-----------|-----------------|------------|--------|
| Proximate | Low             | VS         | VW     |
| Proximate | Med             | Small      | W      |
| Proximate | Adv             | Med        | Med    |
| Moderate  | Low             | Small      | W      |
| Moderate  | Med             | Med. Small | Med    |
| Moderate  | Adv             | Med        | HM     |
| Far       | Low             | Med. Small | Str    |
| Far       | Med             | Lar        | HM     |
| Far       | Adv             | V. Lar     | Str    |

From the table 3.1, it can be observed that if the DS to BS is far and residual energy is advance, then the radius will be very large. If distance to the BS is close and residual energy is low, then the radius became very small. When Distance to the BS is proximate and residual energy is advance, then CH Chance became the medium, and if Distance to BS is far, residual energy is low, then the CH chance becomes the strong.

3.3.5. Inference

In the proposed scheme, the two inputs distance to BS and residual energy are first fuzzed by their membership function, and the type-2 fuzzy is inferred by the rule base, which is shown in Table 3.1. Therefore, the distance to BS and residual energy membership intervals are shown below respectively:

$$[\underline{\mu}_{DB}(t), \overline{\mu}_{DB}(t)] \text{ and } [\underline{\mu}_{RE}(t), \overline{\mu}_{RE}(t)]$$

So, the inference method is as follows:

$$\begin{cases} \overline{f}(t) = \overline{\mu}_{DB}(t) \times \overline{\mu}_{RE}(t) \\ \underline{f}(t) = \underline{\mu}_{DB}(t) \times \underline{\mu}_{RE}(t) \end{cases}$$

(3.5)

where $\underline{f}(t)$ denotes the lower boundaries and $\overline{f}(t)$ express the and upper boundaries.

### 3.3.6. Type Reduction

In this subsection, we find the total crisp output for the inference process while reducing the aggregate type-2 Fuzzy logic into interval type-1 fuzzy logic with the help of the following equation 3.6 and equation 3.7

$$A_1 = \frac{\sum_{n=1}^{L} \overline{f}(t)W(t) + \sum_{t=L+1}^{K} \underline{f}(t)W(t)}{\sum_{t=1}^{L} \overline{f}(t) + \sum_{t=L+1}^{K} \underline{f}(t)}$$

(3.6)

$$A_2 = \frac{\sum_{t=1}^{R} \underline{f}(t)W(t) + \sum_{t=R+1}^{K} \overline{f}(t)W(t)}{\sum_{t=1}^{R} \underline{f}(t) + \sum_{t=R+1}^{K} \overline{f}(t)}$$

(3.7)

where K indicates the range of the sample, $W(t)$ represents the series of weights of the output obtained by the rule resulting from the base rule in increasing order. The left and right limits of the reduced fuzzy sets are seen in A1 and A2. L and R are determined according to the KM algorithm [50].

3.3.7. Defuzzification

From the following equation 8 we can calculate the defuzzification output.

$$A = \frac{A_1 + A_2}{2}$$

(3.8)

## 3.4 Cluster Head Election

For the data transmission, we design a clustering algorithm where the sensor nodes join a cluster and send their data to the CH of the cluster which then forwards it to BS through other CHs in multi-hop transmission. The steps of proposed CH selection algorithm are as follows.

### 3.4.1. Distance Measurement

Before stating the data transmission, each node in the network calculates its distance from BS based on received signal strength indication (RSSI) value of the received start-up packets from BS. As the BS and the nodes are static after installation, the procedure needs to be done only once.

### 3.4.2. Pretendent CH Selection

At the beginning of each round, as each node has the same amount of residual energy, a random probability value, $\mu$ between 0 to 1 is assign to each sensor node. This $\mu$ value is compared with a predefined threshold probability (T). Each node having the $\mu$ greater than T are selected as pretended CH and rest nodes are considered as member sensor nodes.

### 3.4.3. Calculation of Competition Radius

After selecting pretendent CHs, for each pretendent CH, its distance from BS, and residual energy is calculated and used as input variables of fuzzy type-2 logic.
   After applying the fuzzy rule, the competition radius of each pretendent CH is generated.

### 3.4.4. Calculation of CH Chance

Similar to step 3, a CH chance is also generated using fuzzy rules.

### 3.4.5. Selection of Final CH

After the generation of competition radius and CH chance, each pretendent CH broadcast a competition message containing its node Id and CH chance to other pretended CHs inside its competition range. Thus, each pretended CH may receive one or more competition CH message. Each pretendent CH compares its CH chance with the CH chance of each received message, If the CH chance of this CH is greater than all CH chances then it becomes the final CH otherwise it changes its status from pretended CH to member node.

### 3.4.6. Assignment of Nodes to Final CHs

After the selection of the final CHs, the CHs send a control message in their communication range. Each member node that receives only one control message send a join CH message to the sender final CH. In case of a member node receives more than one control messages, it compares its distance from each sender final CH and send a join CH message to the nearest final CH. In worst case if a member node doesn't receive any control message then it declares itself a final CH. This process continues until all sensor nodes join their respective final CHs.

### 3.4.7. Data Transmission

Finally, the CHs send a data message containing time slot tables to the member nodes used to allocate data gathering time slots to each member node. The member nodes send raw data to their respective final CHs according to the slot table.
For data transmission, the CHs close to the BS transmit all data directly to the BS, and the CHs, far from the BS, sends data through a multi-hop process.

### 3.5 Simulation and Results

In this section, we assumed some assumptions for simulations which are shown in table 3.2. Fig. 3.2 shows the energy consumption of TTDFP, CHCCF and proposed algorithm at 1000 rounds. From the figure we can observe that our proposed scheme consumes less energy in comparison to TTDFP and CHCCF algorithms. It also reveals that in the proposed scheme, the number of live nodes is much greater than that of compared algorithms.

Table 3.2 Assumed Parameters

| Parameters | Values |
| --- | --- |
| BS location | Center of the area |
| Node located area (m$^2$) | 100*100 |
| Node location | Randomly distributed |
| No. of deployed nodes | 100 |
| Initial energy | 1 J |
| $\varepsilon_{amp}$ | 0.0010 pJ/bit/m$^4$ |
| $E_{elec}$ | 50 nJ/bit |
| $E_{DA}$ | 5 nJ/bit |

Fig. 3.3 shows the number of dead nodes at every round in the network. From the figure we can see that the number of dead nodes in our proposed algorithm in each round is less than the other compared algorithms. The main reason behind is that in multi-hop transmission the CH require less energy to forward the data to other CH compared to directly sending to the BS. It also shows that our proposed scheme is more robust than the compared algorithms over a longer time.

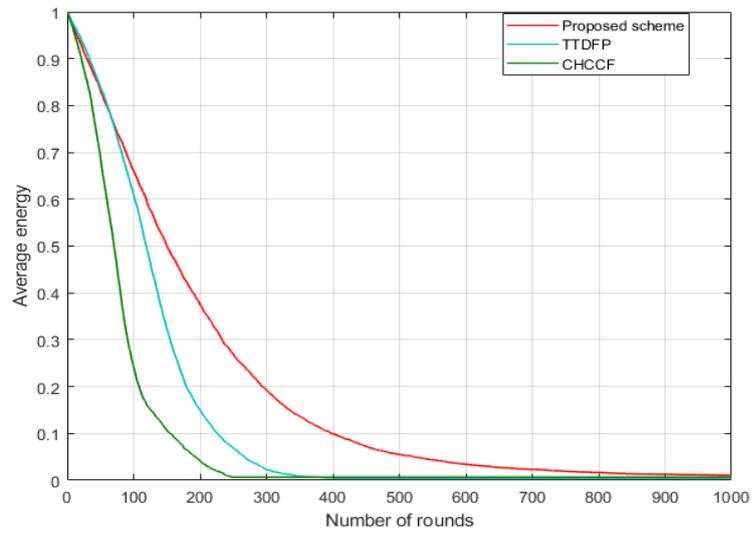

Fig 3.2 No. of rounds vs average energy

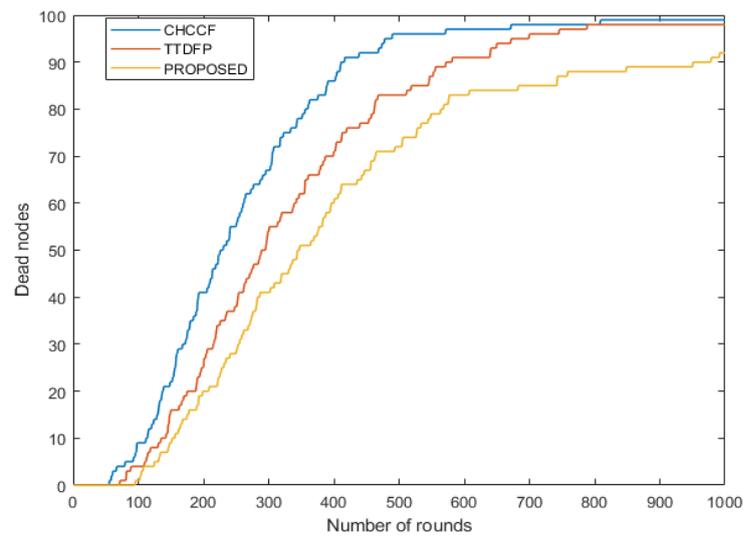

Fig 3.3 No. of rounds vs dead nodes

Fig. 3.4 and Fig. 3.5 demonstrate that the energy efficiency of first node died (FND) and mid or half node died (HND) in the network. we can see from the figure that, the energy efficiency of FND and HND is higher than that of the other compared algorithms. It appears that in the suggested scheme, network durability is better. After the calculation we find that the proposed scheme has better performance than TTDFP by 171.42% and CHCCF by 133.33% in case of FND. In HND the proposed scheme has better performance than TTDFP by 152.21% and CHCCF by 119.03%. Therefore, it can then be observed from all the figures that, for each parameter, the performance of the suggested method is better.

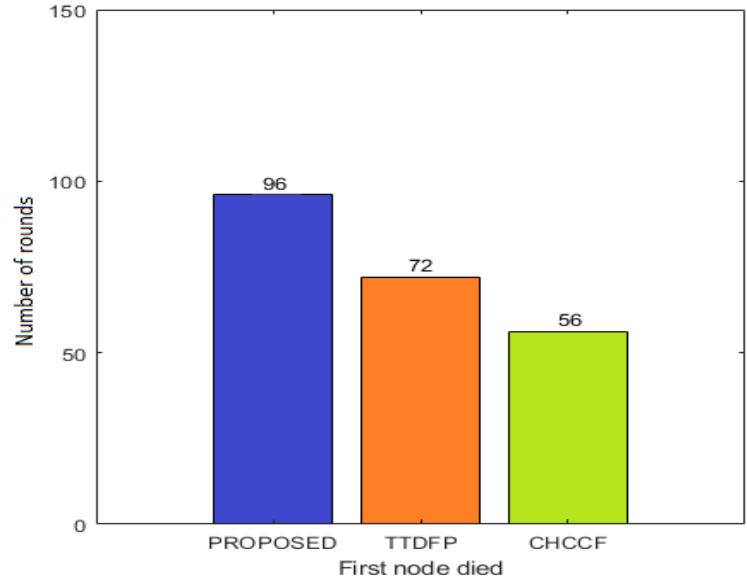

Fig 3.4 FND vs no. of rounds

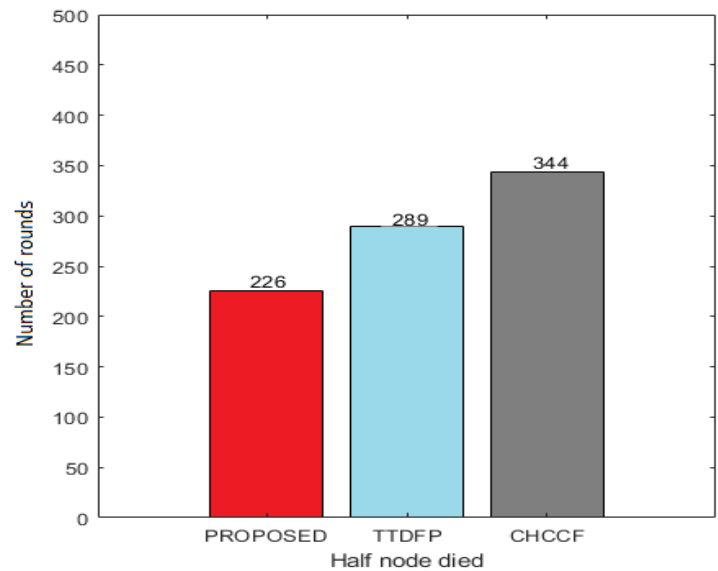

Fig 3.5 HND vs no. of rounds

## 3.6 Chapter Conclusion

In this chapter, we proposed a new scheme for stable clustering in WSN. The issue of hot spots prompted the use of multi-hop routing in a clustering strategy. To fix the issue, we use an energy-efficient, unequal clustering protocol to balance energy consumption between cluster heads (CHs). After that, we add type-2 fuzzy logic to choose the CHs for the whole network. According to the analysis, the proposed system has increased efficiency relative to current clustering methods. It has been shown that the proposed solution is practical and increases the scalability of the network such that the existence of the network greatly improves. The proposed method works for WSNs with stationary homogeneous nodes, but in future work, we will broaden our approach to mobile ad hoc networks with heterogeneous nodes.

# Chapter 4 Utility of Game Theory in Defensive WSNs

## 4.1. Chapter Overview

Wireless Sensor Networks (WSNs) are more vulnerable to computer security threats than traditional networks. Due to the ad-hoc features of WSNs, its nodes are flexible enough to enter or leave a connection leading to different network topologies [50]. Thus, there's no defined formulation for replicating records, whilst the nodes come to be ambiguous. The complexity of the nodes and the inability to define the mode of data replication can lead to problems if a malicious intruder attacks the system. WSNs also experience the problem of limited power, making them behave selfishly to conserve energy, leading to risks of network malfunctioning [51]. Such WSN's limitations make their security schemes more vulnerable to attacks. As a result, research on the security of WSN has increased, and several techniques have come upto mitigate the threats WSNs encounter [52]. The chapter, therefore, critiques the game theory technique as one in all WSN safety strategies. Game principle is a modern-day smart optimization scheme that handles problems wherein fee features of various entities have mutual dependency [53].

The theory has been useful in modeling the behavior of several applications like WSNs. Recent distributed and infrastructure-less system have benefited from the concept of game theory, including decentralized communication systems like WSNs. The model includes the interaction among defenders and attackers mapped to a set of gamers, and each player fights to their benefits. Game theory applies to such a security model involving the action of attackers and defenders. Therefore, the chapter explains the game theory approach in mitigating the WSN threats. The important strategies in the Game theory concept cooperative games and non- cooperative games, attacks, the best protection game method, and the form of games. Apart from cooperative and non-cooperative community environments, the studies additionally evaluations inner and external attack eventualities. In the end, the study proposes possible future research trends, including the possibility of mitigating intelligent attacks using the game theory approach.

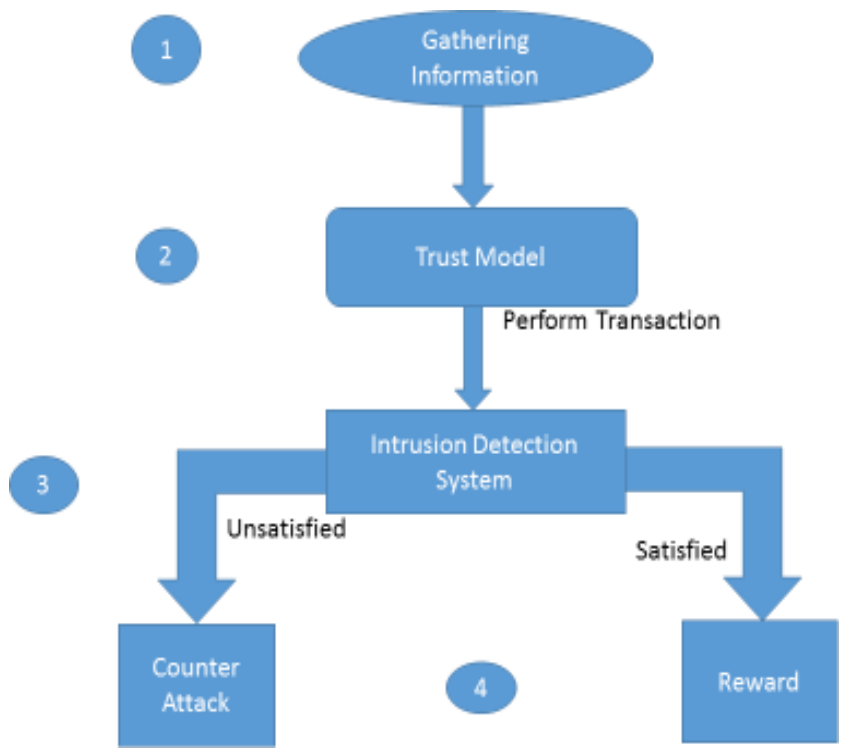

Fig. 4.1 Reputation and General Trust Model Mechanisms

The regular method of a receiving as actual with version is tested in determine 1, this model is much like the version soak up [54]. The popularity and significant trust model is split into four steps. The first step is gathering statistics from the website, site visitor's movement, then the second one step, that applies the fine believe version. The 0.33 step is intrusion detection system through the usage of this step analyzed information at a few degrees inside the accept as true with version. The fourth and final step is in charge of punishing or profitable the infected or generous packets. This latest model aims at accomplishing power green networks towards the intrusion effect the use of the overall principle of gaining knowledge of automaton by using the usage of sampling the incoming packets. The equal precept can be used to boom the protection of WSNs the usage of undertaking concept.

## 4.2. Related Works

Apart from enhancing the safety of WSNs, researchers have advanced efforts to enhance different functions of the device like a battery. The goal is to extend the life of WSNs to the use of power vehicles that fill up the strength to lifetime-essential sensors [55]. Traditional WSNs used energy-confined batteries, therefore, restricting the life of sensor networks. As a result, there have been studies in the past, focusing on the problem of energy replenishment for sensors. Two sensor charging categories have, therefore, come up. One of them enables sensors to collect energy within its surroundings, including solar energy, vibration energy, and wind energy [56]. The varying environmental conditions have, however, limited the performance of the energy harvesting strategy.

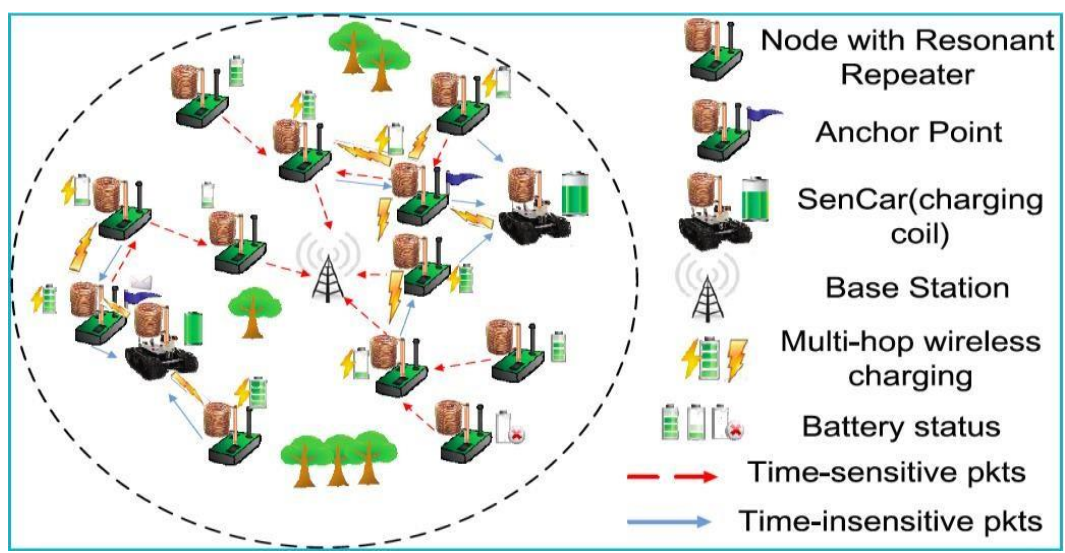

Fig. 4.2 Improve charging functionality for wireless rechargeable sensor networks using resonant repeaters

The second strategy is the employing of mobile charging vehicles that travel to the location of the wireless sensors to charge them. Such vehicles use wireless energy transfer methods to charge the WSN batteries [57].

Charging the wireless sensors through charging vehicles is more reliable than the energy harvesting method because it eliminates the problem of fluctuating environmental conditions. The limitation of the second technique, which assumes that either one or multiple vehicles conduct the charging process. As a result, some sensors which have depleted their power can consume them before replenishing other sensors [58]. Energy charging vehicles have restricted charging potential ranging among 30 and 80 mins. Shown in 4.1

$$s_1, s_2, s_3 \ldots s_6 \tag{4.1}$$

The charging vehicle will first charge the first sensors beginning with *s1*. If the vehicle depletes its energy before reaching *s3*, then *s4, s5, s6* will be uncharged. Shown in 4.2

$$s_4, s_5, s_6 \text{ (uncharged)} \tag{4.2}$$

Just like within the protection of WSNs, studies continue to be underway to enhance the charging capacities of automobiles and the deployment of a couple of charging vehicles. Source: [7]

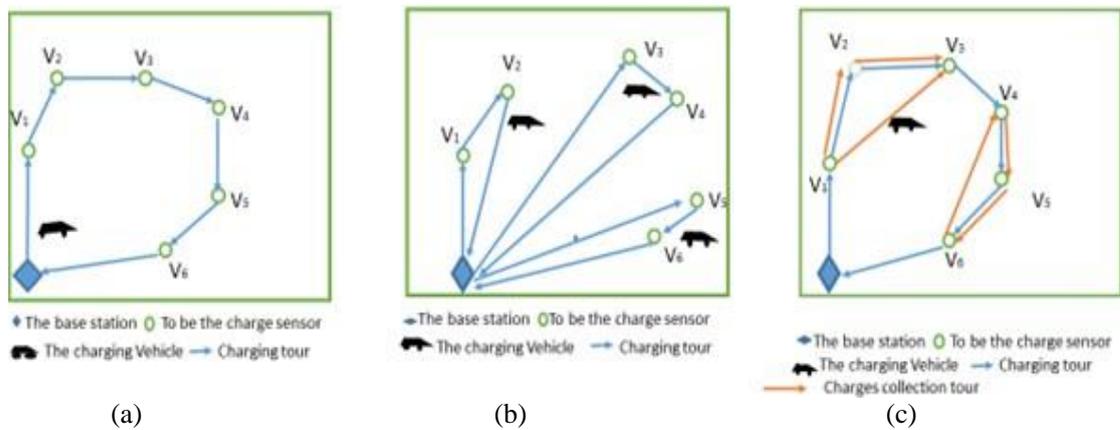

(a)　　　　　　　　　　　　(b)　　　　　　　　　　　　(c)
(a) Employ just one charging vehicle to charge sensor     (b) Dispatch multiple charging vehicle to charge sensors     (c) Deploy one charging vehicle carrying multiple low-cost removable charges to replenish sensors

Fig. 4.3 Charging mechanism of vehicles

## 4.3. Proposed Method

The game theory concept is a department of sensible optimization and represents a game related to player companies that behave cooperatively and non-cooperatively. The player agencies intention to sell their blessings (payoffs) through techniques that different cumulative player movements execute [59]. Game theory has basic definitions of parameters, including:

- Game theory is a strategic interplay between cooperating and non-cooperating pastimes to constraints and the payoff for movements.
- A game is an entity with a described set of gamers N, which under takes rational

- movements Bi, for each participant i. B participant may be a person, a set of people, or a device within the game.
- The utility/playoff is the earnings or loss to each participant for a given action inside the game UI: B→ T, for each participant i as a result of all gamers B = χi ∈ NBi, wherein ∈ is the Cartesian product
- A strategy is an action plan within the game that the players undertake during game play. A strategic game is *(N, (B), (UI))*.
- Nash Equilibrium (NE) is an action optimization profile $b^* \in A$ which does no longer permits any participant I ∈ N to benefit by using deviating from its strategy and picking another action [12][13]. Utility function translates the concept as *UI ($b^*i$, $b^*-i$) ≥ UI ($bi$, $b^*-i$),* for all $bi \in B$. $bi$ is the strategy for player *i* and $b^*-i$ are strategies for all players apart from player *i.*

Application of Game Theory enables in preventing outside intruders and detecting malicious nodes that behave selfishly and overburden the whole community system. For example, the Nash Equilibrium has proved to be a beneficial concept for solving social issues in WSN security. The gaming version for egocentric behavior on sensor nodes claims that WSN offers network services via cooperation many of the nodes. In the process, the nodes consume storage space, consume energy, and other resources. As a result, some nodes will become selfish and non-cooperative due to limited resources, hence, affects the entire network performance. Fig. 4.4 shows the graphical representation of the Game theory and defense against wireless attacks.

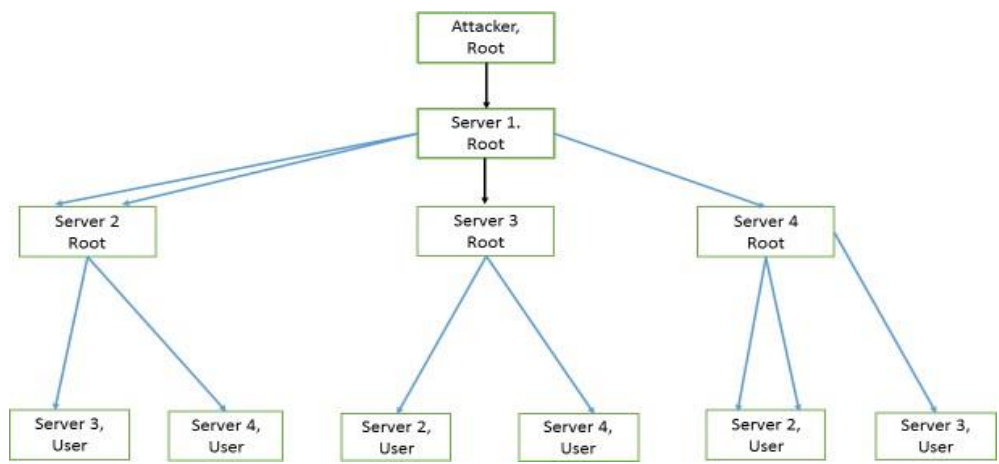

Fig. 4.4 Graphical Representation of the Game Theory and defense against wireless attacks

## 4.4. Proposed Model

Fig. 4.5 Shows the information drift of the general agrees with version. In this version, the selected game is used to face the desiccated assault based on the observed records from the complete community. The seize records are analyzed to pinpoint the community features, assault kind, and appropriate protection strategy to be able to be carried out.

In this model we can see that there are two methods are available in apply defense methods based on the sport and assault. The first manner is cooperative. In this cooperative WSN, nodes calculate three primary parameters: cooperation, security stage and popularity and it calculates for every participant node. These parameters are listed in an amassing factin to neighbor list K(i). After that, each node is checked, if nodes act selfishly or maliciously, according to the behavior of the node is rewarded or punished. If the node is rewarded (benevolent node), the NE lifestyles are checked. A timer is used as a threshold for the NE lifestyles to reduce the risk in the actual-time sensitive applications. This day out is an application established. Then, if NE exists, the highest quality answer is done, in any other case, the node might be implemented on cemoreto the precise defense strategy based totally on game concept. Second, if the node is selfish or malicious, it is going to be seen to a punishment take a look at. Consequently, if the first time, the node acts selfishly or because of hardware damage down a motive, a day trip can be given to the node then the node popularity will go back once more to the neighbor listing K(i). After that, if the node recursively behaves selfishly, the node turns malicious, or the day out expires, the node might be kept out from routing. The above is accomplished until achieving the NE primarily based on the software sensitivity main to the most beneficial answer. On the alternative hand within the non-cooperative, the ranges are offered to begin with an appropriate non-cooperative recreation in opposition to the prevailing attack. Apparently, the same steps are elected as in a cooperative manner however it start from the check step that determines whether the node is benevolent, malicious, or egocentric.

The non-cooperative procedure have two steps. The first is calculating the popularity, cooperation, and security level parameters and the second one step is accumulating the neighbor list.

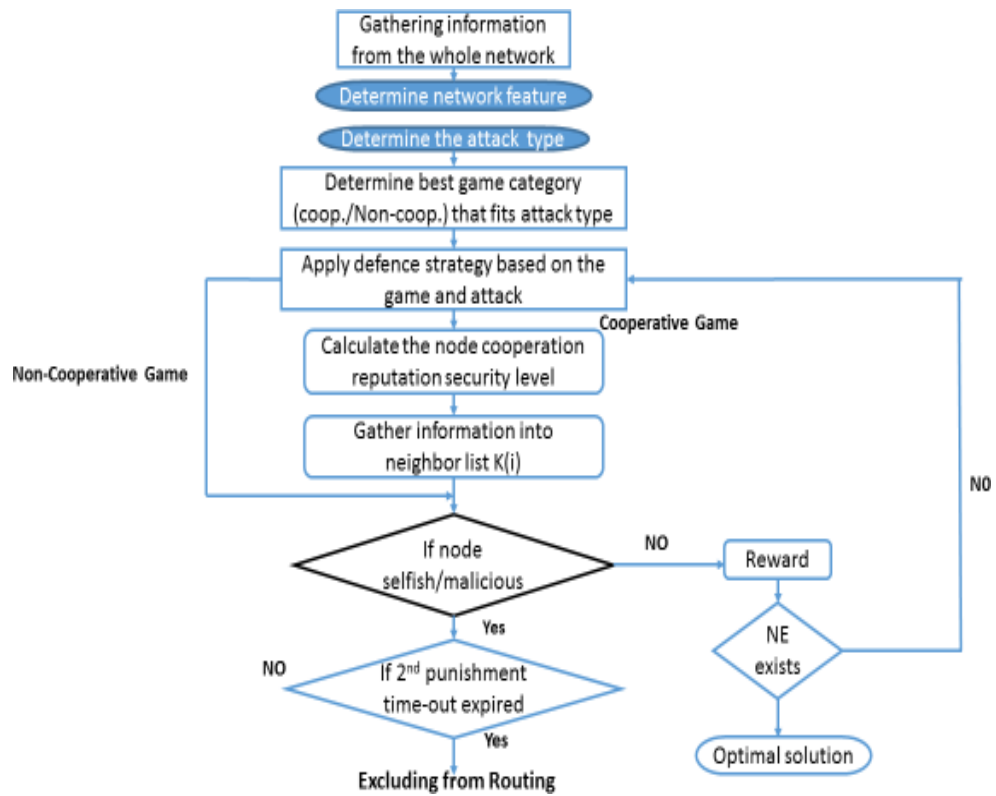

Fig.4.5 Algorithm Proposed Game Theory Trust Model

## 4.5. Proposed Results

This chapter develops a method to evaluate the risk strategy by simulating cooperation among nodes. As a result, the study designed a Trust Derivation Dilemma Game (TDDG) to reduce the overhead of the network. The outcomes of the simulations indicated that the routing scheme could facilitate energy consumption and latency concurrently. In the simulation, the observe distributed numerous sink nodes and several sensor nodes within the community area. Nodes inside the network could sense an event and send the facts packets in an quit-to-end communication channel through a multi-hop. The simulation assumed that the sink nodes were more trust worthy than the standard sensor nodes. The routing scheme tackled active and passive attacks. Sensor nodes in passive attacks gathered sensitive information and became selfish by failing to cooperate with other sensor nodes within the routing process. In active attacks, the malicious nodes requested sensitive information directly. The design of each active and passive attacks can intrude with the normal operations of the Wireless Sensor Networks.

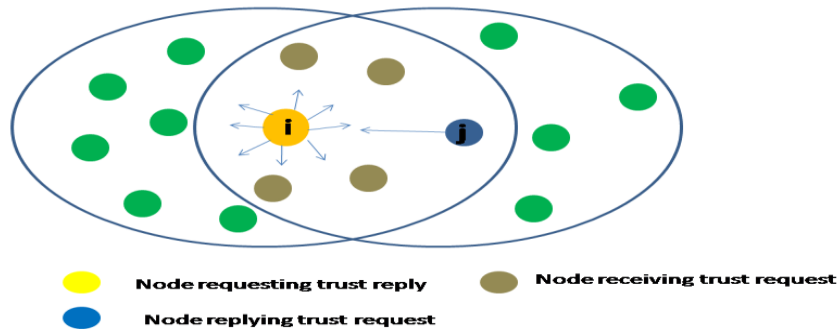

Fig. 4.6 Trust deviation for an arbitrary node

Each sensor node may want to look at the behavior of different nodes within its surrounding or radio range. The results through observation were useful in calculating the trust between the nodes. A random node within the network had $T$ due to direct trust $T_D$ and indirect trust $T_I$. Secure routing and access control both are obtained through the final trust computation model.

In the figure 4.6. node *j* being evaluated and *node i* conducting the evaluation process. A source node *i* initialized the trust deviation process while *j* is target node. As a result, the node *i* broadcasts a trust request packet in the direction of *j*. A hop-limit value will then handle the overhead of the flooding mechanism. A node will evaluate if it had received a similar request as soon as it a trust request packet. Upon confirmation of the reception of the packet, it discards the packet immediately but rebroadcasts the packet if the hop-limit is more significant than zero *(hop-limit>0)*. The other node that receives the packets will check if *j* is its neighbor to unicast a trust reply to source *i* through are verse route [19]. After receiving the advice, the source node *i* computes the trust cost by way of combining each direct and oblique believe. The computed consequences suggest the allows source node i to decide if node j may be relied on or now not. The routing scheme considers a recommendation from neighbors of the nodes that have been evaluated. The protocol values neighbors' recommendations because the neighbors can accurately observe the malicious behavior of other nodes and reduce network overhead. The success of the simulation depends on an efficient design, which ensures that the trust information from the neighbors fulfils the security requirements [20][21]. The TDDG design had *N* nodes, which are the total number of evaluated neighbors' nodes, excluding the evaluating nodes. Two strategies for participating nodes include 'reply' and 'no-reply' whereby the former indicates the events to which the nodes replied to the evaluating node during the trust request. 'No reply' refers to nodes that ignored the trust request even after receiving it.

## 4.6. Chapter Conclusion

The results from this study indicate the in most cases, executing the game theory in a WSN helps reduce the network overhead in its communication channel. By evaluating the response to the trust request, it becomes easier to eliminate non-cooperative nodes while maintaining the cooperative ones. The study has used Game Theory as an efficient approach towards WSN but its real-life application is still a challenge and therefore requires further studies. Researchers have applied efforts to study simple energy saving mechanisms and topology control applications. However, there needs to be more research to improve the theory in explaining moving arget tracking, scheduling sensing tasks, QoS control, and WSN security under distributed conditions.

# Chapter 5 Conclusions and Future works

## 5.1. Conclusions

WSNs are vulnerable to a variety of security threats that can negatively impact of application success. Security in WSNs is difficult due to limited energy, communication bandwidth, and computational power. Furthermore, sensors are frequently deployed in an open environment with no physical security. Given the range of WSN applications and possibly varying security requirements, we believe that special enhancing security approaches to securing WSN are required. This dissertation envisions three key open challenges: a) Fuzzy based unequal clustering with multi-hop communication for wireless sensor networks. b) Type-2 fuzzy logic-based cluster head selection in wireless sensor networks. c) Application of game theory in defensive wireless sensor networks. This dissertation solves one or more problems by proposing fuzzy interference system algorithm and theoretical framework for wireless sensor networks related scenarios and present simulation studies to establish the efficacy of the proposal.

Routing is a major part of data aggregation and providing an optimized and energy efficient routing scheme is one of the major open challenges for wireless sensor networks. Moreover, it is observed that the literature lacks a unified approach to wireless sensor networks data routing that supports stationary WSN. Chapter 2 of this dissertation proposed an unequal clustering scheme for wireless sensor networks based on fuzzy logic which enhance the network lifetime. Later this chapter using multi-hop data transmission for reducing the energy consumption. Also, this chapter simulates different scenarios to prove the proposed work accurate.

The Salability problem in wireless sensor networks is a critical issue due to the extremely high node numbers and relatively high node density. A strong routing protocol should be scalable as well as adaptable to changes in network topology. Therefore, when the network size or workload increases protocol must perform better. This problem is considered in chapter 3. This chapter proposed type -2 fuzzy based cluster head selection in WSNs. This chapter initially using multi-hop technique to reduce energy consumption later using interval type-2 fuzzy logic chooses cluster head to reduce the scalability problem. As a result, network lifetime increases.

Game theory is a viable mathematical tool that models interactive decision-making processes and can be widely distributed in wireless sensor networks. Because of its exceptionalism in regulating the behaviors of many players, game theory is widely used in the energy-efficient algorithm for WSNs. The chapter 4 discusses the application of game theory approach in mitigating the WSN threats. Initially this chapter demonstrates the uses of game theory in WSN to prevent different type of attacks such as selfish nodes and malicious behavior. The study explains the defense strategies of WSN from both internal and external attacks using game theory scheme. In addition, the chapter emphasizes the general trust model for decision making that employs the game theory framework. Furthermore, the article illustrates the theory's importance in ensuring WSN security from acute attacks, as well as its role in improving data trustworthiness and node cooperation in various WSN architectures.

## 5.2. Future works

The dissertation has demonstrated that energy consumption and network life time is the main issue in the wireless sensor networks, so it is crucial to control the performance of the network. In order to prevent the network from different types of issues, there are several exciting extensions to our work. The interval type-2 fuzzy logic theory to solve the secure clustering protocol for mobile ad-hoc networks and the light weight secure mechanism in chapter 2 can be further extended by taking consideration of more parameters such as distance to base station, residual energy, and node density. Moreover, the scenario can be formulated into an optimization problem with constraints depending on the application scenario. In chapter 3, unequal clustering problem can be solved using three type 2 fuzzy input parameters and also taking multi-hop technique for data transmissions to make the network stable in WSN. In chapter 4 detecting and preventing the WSN from bad-mouth attacks strategy can be extended.

First by using evidence method detect the attacks in the nodes and prevents the node from bad-mouth attacks using game theory approach.

Thesis for Master's Degree of CQUPT                                                               References[54] F.G.Mármol, G.M.Pérez, "Trmsim-WSN, Trust and Reputation Models Simulator for Wireless Sensor Networks," *In Proceedings of the International Conference on Communications (ICC),* Dresden, Germany, pp. 1–5, 14–18 June 2009.

[55] Tao Zou, Wenzheng Xu, Weifa Liang, Jian Peng, Yiqiao Cai, Tian Wang, " Improving the Charging Capacity for Wireless Sensor Networks by Deploying One Mobile Vehicle with Multiple Removable Chargers," *Ad Hoc Networks*, 63, 79-90, (2017).

[56] Aman Kansal, J.Hsu, M.Srivastava, & V.Raghunathan, "Harvesting Aware Power Management for Sensor Networks," *Proc. 43rd ACM Annu. Design Automat*, 651-656, 2006.

[57] Cong Wang, Ji Li, Fan Ye and Yuanyuan Yang, "Improve Charging Capability for Wireless Rechargeable Sensor Networks Using Resonant Repeaters," Proc. IEEE 35th, *Int. Conf. Distributed Computer Systems*. (ICDCS), 133-142, 2015

[58] Wenzheng Xu, Weifa Liang, Xiaola Lin, Guoqiang Mao, "Efficient Scheduling of Multiple Mobile Chargers for Wireless Sensor Networks," *IEEE Trans. Veh. Technology,* 65(9), 7670-7683, 2016.

[59] Beibei Wang, Yongle Wu & K.J. Ray Liu, "Game Theory for Cognitive Radio Networks: An Overview," *Journal of Computer Networks*, 54, 2537-2561, 2010.
70

# Acknowledgement

First and foremost, I am extremely thankful to my parents Niyaz Ahmad and Noorun Nishan for their love, prayers, caring, and sacrifices for educating and preparing me for my future. Also express my thanks to my sister Neda Tabassum and my lovely brothers Mohd Arshad and Mohd Faizan for their love, understanding, prayers and continuing support to complete this research work.

Next I would like to express my deep and sincere gratitude to my research supervisor Dr. Yang Tao for giving me the opportunity to do research and providing invaluable guidance throughout this research. His dynamism, vision, sincerity and motivation have deeply inspired me. He has taught me the methodology to carry out the research and to present the research works as clearly as possible. It was a great privilege and honor to work and study under his guidance. I am extremely grateful for what he has offered me. I would also like to thank him for his friendship, empathy, and great sense of humor. I am extending my heartfelt thanks to his wife, family for their acceptance and patience during the discussion I had with him on research work and thesis preparation.

Finally I would like to thank to my co-supervisor Dr. Liu Yang and my friend Tazeem Ahmad who helped me to cope up with a new academic and living environment. Most importantly I am very much grateful to the Chinese government scholarship council for the generous scholarship to support my education in Chongqing, China.

# Publications and Achievements

- **Mohd Adnan,** TaoYang,, Tazeem Ahmad, Liu yang, "An Unequally Clustered Multi-Shop Routing Protocol based on Fuzzy Logic for Wireless Sensor Networks." **IEEE ACCESS JOURNAL** DOI: 10.1109/ACCESS.2021.3063097.

- **Mohd Adnan**, Tao Yang, S.k. DAS, Tazeem Ahmad, " Utility of game theory in defensive Wireless Sensor Networks (WSN) from selfish nodes and malicious behavior" **3RD INTERNATIONAL CONFERENCE ON INNOVATIVE COMPUTING ANDCOMMUNICATION(ICICC-2020)** DOI: 10.1007/978-981-15-5113-0_75.

- **Mohd Adnan,** Tazeem Ahmad, Tao Yang, "Type-2 Fuzzy Logic Based Energy-Efficient Cluster Head Election for Multi-Hop Wireless Sensor Networks." (**IEEE APWiMob 2021**) (Presented).

- Got one of the prestigious fully-funded Chinese Government Scholarship (2018-2021).

- Got first prize in Essay Writing Competition organized by CQUPT China in 2019.